\documentclass[preprint]{revtex4-1}
\usepackage[utf8]{inputenc}
\usepackage{amsmath}
\usepackage{amsfonts}
\usepackage{amssymb}
\usepackage{makeidx}
\usepackage{graphicx}
\usepackage{epsfig}
\usepackage{array}
\usepackage{tikz}
\usepackage{multirow}
\usepackage{xcolor}
\usepackage{epstopdf}
\usepackage[colorinlistoftodos]{todonotes}
\usepackage[compat=1.1.0]{tikz-feynman}
\usepackage[export]{adjustbox}
\begin{document}
\title{ Dark matter mass from relic abundance, an extra $U(1)$ gauge boson,  and  active-sterile neutrino mixing}
\author{\bf{Imtiyaz Ahmad Bhat}}
\email{imtiyaz@ctp-jamia.res.in}
\affiliation{Centre for Theoretical Physics, Jamia Millia Islamia (Central 
			University), Jamia Nagar, New Delhi 110025, India }
\author{\bf{Rathin Adhikari}}
\email{rathin@ctp-jamia.res.in}
\affiliation{Centre for Theoretical Physics, Jamia Millia Islamia (Central University), Jamia Nagar, New Delhi 110025, India }
\begin{abstract}
In a model with  an extra $U(1)$ gauge  to SM gauge group, we have shown the  allowed region of masses of extra  gauge boson  and the dark matter  which is the lightest one among other right-handed Majorana fermions present in the model. To obtain this region, we have used bounds coming from constraints on active-sterile neutrino masses and mixing from various oscillation experiments,  constraint on dark matter relic density obtained by PLANCK together with the constraint on   the extra gauge boson mass and its gauge coupling recently obtained by ATLAS Collaboration at LHC. From the allowed regions, it is possible to get some lower bounds on the masses of the extra gauge boson and the dark matter and considering those values it is possible to infer what could be the spontaneous symmetry breaking scale of an extra $U(1)$ gauge symmetry.
\end{abstract}
\maketitle
 \section{Introduction}
 Although Standard Model (SM) has got tremendous success in describing various phenomena at the  elementary particle level, but SM failed to account for two major experimental results, one related to the existence of dark matter (DM) \cite{Bertone:2004pz} in the universe and the other related to neutrino oscillation phenomena that require  neutrinos to be massive  and  significant mixings among different flavors of neutrinos. To accommodate neutrino masses and a viable dark matter candidate something beyond SM is necessary. One such example is minimal extension of SM gauge group with extra $U(1)_{X}$ gauge symmetry. Additional symmetries \cite{symmetry} either global or gauged are imposed which play the role in guaranteeing the stability of dark matter candidate. There are several $U(1)_{X}$ gauge extended models with minimal extension to the SM \cite{Langacker:2008yv,u2,Ma:2006km}. An important feature of these models is that in comparison to SM there is one extra neutral gauge boson. In general there could be mixing of the extra gauge boson $X$ with the SM $Z$ boson, which results in the modification of neutral current phenomena. The $Z$ pole data could be affected indirectly through such mixing and
 could shift the measured $Z$ mass and its coupling to SM fermions. But nice agreement of the mass and coupling with SM predictions constrains such mixing to be  lower than  1$\%$ \cite{mix}.

 On the other hand, in cosmology to explain the rotational curves of the heavy massive body inside the galaxies, one need to propose the presence of dark matter \cite{Zwicky:1933gu,wmap}.  Dark matter relic density has been constrained from PLANCK  experiment \cite{Aghanim:2018eyx}.
 \begin{equation}
 \Omega_{DM} h^2=0.1200\pm0.0012
 \end{equation}
where $\Omega_{DM}$ is the density parameter for dark matter and $h=H_{0}/$(100 km $s^{-1}$ Mp$c^{-1}$. 
Recently CMS and  ATLAS \cite{Aaboud:2017yvp,Aaboud:2017buh,cms} Collaborations at LHC have obtained stringent bound on the mass and gauge coupling associated with the extra $U(1)$ gauge boson. 
In the  light of recent neutrino oscillation phenomena \cite{Gonzalez-Garcia:2015qrr,Ahn:2006zza}, there is a proposition of presence of sterile neutrino apart from three active neutrinos. There are recent indications in the Fermilab experiment \cite{Adamson:2017zcg} about some nonzero mixing among active and sterile neutrinos with sterile neutrino mass in the eV scale \cite{fermi}.
 
In connection with these observational results, we have considered here an $U(1)$ gauge extended model \cite{Adhikari:2015woo}, which contains dark matter fields and also can accommodate active-sterile neutrino masses and mixing. In this model right handed Majorana fermion is found to be suitable candidate for dark matter as discussed later. There are some studies on the constraints 
on model parameters of $U(1)$ gauge extended models based
on collider phenomenology and cosmological constraints \cite{okada}. However, in this work we have shown in detail the allowed region in the dark matter mass $m_\psi$ and extra $U(1)$ gauge boson mass $M_X$  plane. For that we have  considered PLANCK constraint on dark matter relic abundance. Besides, we consider constraints coming from active and sterile neutrino masses and  their mixing, to
find the allowed region. In the
model considered here, the presence of appropriate active and sterile neutrino mass and their mixing requires the presence of appropriate range of mixing (angle $\theta $) of dark matter with another
right handed Majorana fermion and their mass gap $\Delta$ as discussed later. In any extra $U(1)$ gauge model, the Majorana fermion (which is dark matter in our case) can be annihilated to SM fermion and antifermions through $X$ boson as mediator. However, in our work, due to active sterile mixing resulting in  the nonzero value of $\theta$, the co-annihilation of dark matter with other Majorana fermion is also present. So the observed dark matter relic density will depend on both annihilation and co-annihilation of dark matter in general in our work. With LHC constraint along with relic abundance constraint and constraint on $\Delta$ and $\theta$ (from neutrino oscillation data) we have also studied the possibility of lower bounds on mass of $X$ boson  and the  dark matter mass and the corresponding $U(1)$ gauge coupling $g_X$. 

Particularly the allowed region is more for higher values of $M_X$ and $m_\psi$. It is found that the allowed regions does not significantly vary with $\Delta$ values but more sensitive to the variations of $\theta$ - the mixing of the dark matter with other heavy right handed Majorana field considered in the model. All these analysis has been done considering $Z-X$ mixing to be zero at the tree level. Later on we have shown that higher order corrections of this $Z-X$ mixing remains very small of the order of $10^{-5}$ and has been neglected.
 
 In section II, we have discussed the salient features of a model which is $U(1)$ extension of the SM gauge group and the model can successfully explain active and sterile neutrino mass and  mixing and also there is scope of dark matter. We have discussed the interaction of right-handed Majorana field in the mass basis with extra gauge boson which will be useful for calculation of cross section for annihilation and co-annihilation of dark matter. In section III, using the experimental data on the active and sterile neutrino mass and their mixing, we have obtained the allowed region of  the mass difference parameter $\Delta$ of dark matter with the next heavier right-handed Majorana fermion and their mixing angle $\theta$.
  In section IV, dark matter relic density has been studied, taking into account annihilation as well as co-annihilation of dark matter with next heavier right-handed Majorana fermion going into final states of SM fermion antifermion pair. In section V in finding allowed model parameters, we have considered certain allowed values of $\Delta$ and $\theta$ as obtained in section III from active and sterile neutrino mass and mixing. We have obtained the allowed parameter space for dark matter mass $m_{\psi}$, $X$ boson mass $M_{X}$ based on constraints coming from LHC and relic abundance and also neutrino oscillation mass and mixing constraint corresponding to various  $\Delta$ and $\theta$ values. In section VI we have discussed possible modification of $Z-X$ mixing after including higher order correction. In section VII we have concluded about our work.
  
 \section{The Model}
 Here we have considered a model \cite{Adhikari:2015woo} which is an $U(1)$ extension of SM, in which neutrino masses have been studied extensively and the mass of neutrinos has been connected to dark matter which is stabilized by a residual $Z_2$ symmetry of the spontaneously broken $U(1)$ gauge symmetry. The model has only one electro-weak symmetry breaking doublet $\phi^{(+,0)}$ from which  tree level masses to quarks are obtained. Charged lepton masses are generated at one loop level with dark matter as mediator. The three active and one sterile neutrino masses as well as mixing between active-sterile neutrinos have been generated through one-loop. The model contains two heavy right handed fermion triplet $\Sigma^{(+,0,-)}_{1R,2R}$ and three neutral singlet fermions $N_R, S_{1R}, S_{2R}$. These have been chosen so as to cancel all anomalies with each other.  $U(1)_X$ gauge symmetry is spontaneously broken by singlet scalar $\chi_{1,2}^0$ and residual $Z_2$ symmetry is obtained.    The other scalars are added to obtain masses for all fermions. There are two scalar doublets $\eta_{1,2}$ - one couple to $S_{1R}$ and other to two $\Sigma$'s.  
 
 \begin{table}[htb] 
\begin{center}
\begin{tabular}{| m{3cm} |m{2.5cm} |m{1cm}|}
\hline
particle &  $U(1)_{X}$ & $Z_2$ \\ 
\hline \hline
$(u,d)_L$& $n_{1}$ & + \\ 
$u_R$ & $\frac{1}{4}(7n_{1}-3n_{4})$ & + \\ 
$d_R$ & $\frac{1}{4}(n_{1}+3n_{4})$  & + \\ 
\hline
$(\nu,l)_L$  &  $n_{4}$ &+ \\ 
$l_R$ & $\frac{1}{4}(-9n_{1}+5n_{4})$  & + \\ 
\hline
$\Sigma^{(+,0,-)}_{1R,2R}$ & $\frac{3}{8}(3n_{1}+n_{4})$& -- \\ 
\hline
$N_{R}$ & $-\frac{3}{4}(3n_{1}+n_{4})$ & + \\ 
\hline 
$S_{1R}$ & $\frac{1}{8}(3n_{1}+n_{4})$ & --  \\ 
$S_{2R}$ & $\frac{5}{8}(3n_{1}+n_{4})$ & -- \\ 
\hline 
\end{tabular}
\label{Table1}	
\end{center}
\caption{Fermion fields in the model }
\end{table}

\begin{table}[htb]
\begin{center}
\begin{tabular}{| m{3cm}  |m{2.5cm} |m{1cm}|}
\hline
particle & $U(1)_{X}$ & $Z_2$ \\ 
\hline \hline
$\phi^{(+,0)}$  & $\frac{3}{4}(n_{1}-n_{4})$& + \\ 
\hline
$\eta^{(+,0)}_1$  & $\frac{1}{8}(3n_{1}-7n_{4})$& -- \\  
$\eta^{(+,0)}_2$ & $\frac{1}{8}(9n_{1}-5n_{4})$ & -- \\ 
\hline
$\chi^0_1$ & $\frac{1}{4}(3n_{1}+n_{4})$ & + \\
$\chi^0_2$ & $\frac{3}{4}(3n_{1}+n_{4})$& + \\
\hline
$\chi^0_3$  & $\frac{1}{8}(3n_{1}+n_{4})$& -- \\ 
$\chi_4^+$ & $\frac{3}{8}(n_{1}-5n_{4})$ & -- \\ 
$\xi^{(++,+,0)}$ & $\frac{1}{8}(9n_{1}-13n_{4})$ & -- \\
\hline 
\end{tabular}
\label{Table2}
\end{center}
\caption{Scalar fields in the model.}
\end{table}

The fermionic and the  scalar particles  of the model are given in Tables I and  II, respectively. Although there are several $U(1)$ charges corresponding to different fields but using the anomaly cancellation equations all of them can be expressed in terms of the other two $U(1)$ charges $n_{1}$ and $n_{4}$,  corresponding to quark doublet and lepton doublet respectively. Under $Z_{2}$ symmetry, odd and even fields are specified in the last column of the above tables.  
The relevant Yukawa Lagrangian part of the model is:
	\begin{align*}
	L\supset h_{ij}^{\Sigma}\Sigma_{iR}\Sigma_{jR}\bar{\chi_2^0} + h_{12}^{S}S_{1R}S_{2R}\bar{\chi_2^0} +h_{11}^{S} S_{1R}S_{1R}\bar{\chi_1^0}+h_{23}^{N} N_RS_{2R}\chi_3^0 +h_{ij}^{\eta_{2}}\bar{\Sigma_{jR}^0}\nu_i\eta_2^0 + h_{i1}^{\eta_1}\bar{S_{1R}}\nu_i\eta_1^0 \\ +h_{ij}^{\eta_{2}} l_{iL}\Sigma_{jR}\eta_{2}^{+}  +    h_{ij}^{\xi} l_{iR}\Sigma_{jR}\xi^+ +   h_{i1}^{\eta_1}l_{iL}S_{1R}\eta_1^+   + h_{i2}^{\chi} l_{iR}S_{2R}\chi_4^+
\end{align*}
 The first six terms are relevant for masses of Majorana fermions  (shown in Table 1) and also these terms are relevant for active-sterile mixing of neutrinos at one-loop level as discussed later. Last four Yukawa interactions are relevant for charged lepton mass generation and have not been discussed here. $\chi_{1,2}^0$  which breaks $U(1)_X$ gauge symmetry spontaneously,  give masses to 
 $\Sigma^{(+,0,-)}_{1R,2R}$ and $ S_{1R}, S_{2R}$ through interaction as shown in first three terms above and so these scalars have even $Z_2$ parity.    Also from these interactions, $U(1)_X$ charge of  $\chi_{1,2}^0$ is specified by the corresponding  $U(1)_X$ charge of triplet and singlet fermions as shown in the Table.  The active neutrinos get masses at one loop level,  through interactions shown in fifth and sixth terms in above interactions Lagrangian \cite{Adhikari:2015woo}.  From $U(1)_X$ charges of the active neutrinos $\nu_i $ and the 
 $ S_{1R}, S_{2R}$ fermions it follows that the scalar fields in fifth and sixth terms is different from $\chi_{1,2}^0$. They do not have non-zero vacuum expectation value ({\it vev}). So $\eta_{1,2}^0$ is considered an odd under $Z_2$. The SM fields are even under $Z_2$. So  from fifth and sixth terms in above interactions, it follows that $\Sigma^{(+,0,-)}_{1R,2R}$ and $ S_{1R}, S_{2R}$ are odd under $Z_2$. Interestingly, this oddness is decided by the  $U(1)_X$ charge  as discussed.
 Thus these fermions could play the role of dark matter. From the required interactions for one loop fermion masses for sterile neutrinos $N_R$ and charged leptons, the $Z_2$ parity of other nonstandard model  fields are decided \cite{Adhikari:2015woo}.
 $N_R$ is the singlet neutrino which is massless at tree level. Mixing of $N_R$ with active neutrinos as shown later in Fig. 1 and also the mass of $N_R$ at one loop level are obtained through interactions of $N_R$ with $S_{2R}$ and a scalar field $\chi_3$. Here also  scalar field different from $\chi_{1,2}$ is required because of the $U(1)_X$ charges of the fermions in this interaction and $\chi_3$ which does not have non-zero {\it vev} is required to be odd under $Z_2$.  Then $N_R$ is required to be even under $Z_2$ and  and for that it is suitable for consideration as light sterile neutrino. The neutral
 scalars, odd under $Z_2$, in their mass basis have components which are in general, not electroweak singlets and as such they are not good dark matter candidate. This is because they will have too large cross section for their direct detection in underground experiments because of their interactions with $Z$ boson. $\Sigma_{1R,2R}^0$ as dark matter has been discussed in \cite{ma1} and they do not play role in active-sterile neutrino mixing. However, $S_{1R,2R}$ could play the role of dark matter and also lead to active and sterile neutrino mixing as mediator at one loop and has been considered as possible dark matter candidate in our work. 
 
We discuss in short the generation of mass of extra gauge boson $X$ and its mixing with SM  neutral gauge boson $Z$. Let  the  vevs of various neutral scalars fields be $\langle\phi^0\rangle=v_{1}$ and $\langle\chi_{1,2}^0\rangle=u_{1,2}$, then the mass-squared elements, that determines mass for $Z$ and $X$, are given as,
\begin{align}
M_{ZZ}^2=&\frac{1}{2}g_{Z}^2\left(v_{1}^2\right)\\
M_{ZX}^2=&M_{XZ}^2=\frac{3}{8}g_{Z}g_{X}\left(n_{1}-n_{4}\right)v_{1}^2\\
M_{XX}^2=&\frac{1}{2}g_{X}^2\left(3n_{1}+n_{4}\right)^{2}\left(u_{1}^2+9u_{2}^2\right)+\frac{9}{8}g_{X}^2\left(n_{1}-n_{4}\right)^2v_{1}^2
\end{align}

Although in general, there is $Z-X $ mixing but it is expected to be very small so that electroweak precision measurements could be satisfied. The condition for no $Z-X$ mixing between neutral electroweak gauge boson and the extra $U(1)_{X}$ gauge boson is obtained for $M_{ZX}^2=0$ which gives $n_{1}=n_{4}$. With this zero mixing consideration, the mass of the extra $U(1)_{X}$ gauge boson is
\begin{align}
M_{XX}^2=\frac{1}{2}g_{X}^2(4n_{1})^2\left(u_{1}^2+9u_{2}^2\right)
\label{massxboson}
\end{align}
Later on, we consider this zero mixing condition in dark matter relic density calculation.

Since the dark matter is Majorana in nature, its vector coupling with $X$ boson is zero and  it has only non-zero axial-vector coupling with  $X$. The vector coupling $g_{fv}$ and axial-vector coupling $g_{fa}$ of the SM fermion fields with an extra gauge boson are given in Table III. These couplings are related to the chiral couplings \cite{Langacker:2008yv} as follows:
\begin{equation}
g_{f(v,a)}=\frac{1}{2}\left[\epsilon_{L}(f)\pm\epsilon_{R}(f)\right]
\label{22}
\end{equation}

\begin{table}[h!]
\centering
 \begin{tabular}{|| c |c |c||} 
 \hline
  & $g_{fv}/g_X$ &$ g_{fa}/g_X$ \\ [1.5ex] 
 \hline\hline
 \textit{l=e},$\mu,\tau$ & $\frac{9}{8}(n_4-n_1)$ & $\frac{1}{8}(n_4-9n_1)$ \\ [1.5ex] 
 \ $\nu_l$ & $\frac{n_4}{2}$& $-\frac{n_4}{2}$\\[1.5ex]
 \textit{U} & $\frac{1}{8}(11n_1-n_4)$ &  $\frac{3}{8}(n_1-n_4)$ \\[1.5ex]
 \textit{D} &  $\frac{1}{8}(5n_1+3n_4)$&  $\frac{3}{9}(n_4-n_1)$ \\[1.5ex]
 \hline\hline
 \end{tabular}
 \label{Table3}
 \caption{Couplings of SM fermions with extra gauge boson X in terms of $U(1)$ charges $n_{1}$ and $n_{4}$. $U$ and $D$ are up and down type quarks respectively}
\end{table}
\noindent
where the chiral couplings $\epsilon_{L,R}(f) $ are  $g_{X}$ times $U(1)_{X}$ charges corresponding to  left- and right-handed chiral fields as shown in Table III.

Let the mass eigenstates of the four Majorana fermions $S_{1R}, S_{2R},\Sigma_{1R}^{0}, \Sigma_{2R}^{0}$ be $\psi_{k}$ with mass $m_{\psi_{k}}$. The interaction basis $\psi_{j}^{'T}=[S_{1R}, S_{2R},\Sigma_{1R}^{0}, \Sigma_{2R}^{0}]$  could be written in terms of this mass basis $\psi_{k}$  as :
\begin{equation}
 \psi_{j}^{'}=\sum_{k}z_{jk}\psi_{k}
\end{equation}
 with $j,k=1,..4$
where $\Sigma_{1R}^{0}$ and $\Sigma_{2R}^{0}$ are $SU(2)_{L}$ triplets and $S_{1R}$ and $S_{2R}$ are  singlets. One of the lightest among $\psi_{k}$ say, $\psi_{1}$ is a dark matter candidate in this model, which we assume that it  mainly contains  $S_{1R}$ and $S_{2R}$.   We consider  $\psi_2$   as the  next to lightest  among these four mass eigenstates and the masses $m_{\psi_1}$ and $m_{\psi_2}$ are not far apart.  

In considering interactions of extra gauge boson X with $S_{1R}$ and $S_{2R}$  in the mass basis of $\psi_{k}$, we are considering for simplicity that $z_{ij}$ mixing matrix elements has non-zero 1-2 block with mixing angle $\theta$ which is decoupled from 3-4 block.  Then the interaction 
can be written as,
\begin{eqnarray}
\sum_{i,j}\bar{S_{iR}}\gamma_{\mu}(g_{ij}\gamma^{5})S_{jR}X_{\mu}&=&(g_{S_{1R}a}\cos^{2}{\theta}+g_{S_{2R}a}\sin^{2}{\theta})\bar{\psi_{{1}}}\gamma_{\mu}\gamma^{5}\psi_{{1}}X_{\mu}\nonumber\\
&&+(g_{S_{1R}a}\sin^{2}{\theta}+g_{S_{2R}a}\cos^{2}{\theta})\bar{\psi_{{2}}}\gamma_{\mu}\gamma^{5}\psi_{{2}}X_{\mu}
\nonumber\\
&&+ \frac{1}{2}\sin{2\theta}(g_{S_{1R}a}-g_{S_{2R}a})\bar{\psi_{{1}}}\gamma_{\mu}\gamma^{5}\psi_{{2}}X_{\mu}
\nonumber\\
&&+\frac{1}{2}\sin{2\theta}(g_{S_{1R}a}-g_{S_{2R}a})\bar{\psi_{{2}}}\gamma_{\mu}\gamma^{5}\psi_{{1}}X_{\mu}
\label{12}
\end{eqnarray}
where  $i,j=1,2$,  $g_{S_{1R}a}=5/8(3n_1+n_4)g_{X}$ and $g_{S_{2R}a}=1/8(3n_1+n_4)g_{X}$. 
Here $g_X$ is the gauge coupling for extra gauge boson and  subscript $a$ denotes that these are axial-vector couplings. The interactions shown in terms of mass basis 
will be useful in section IV in our calculation of cross section of annihilation and co-annihilation of dark matter.

\section{Active and sterile neutrino mass and mixing}
\begin{figure}[h]
 \includegraphics[scale=0.5]{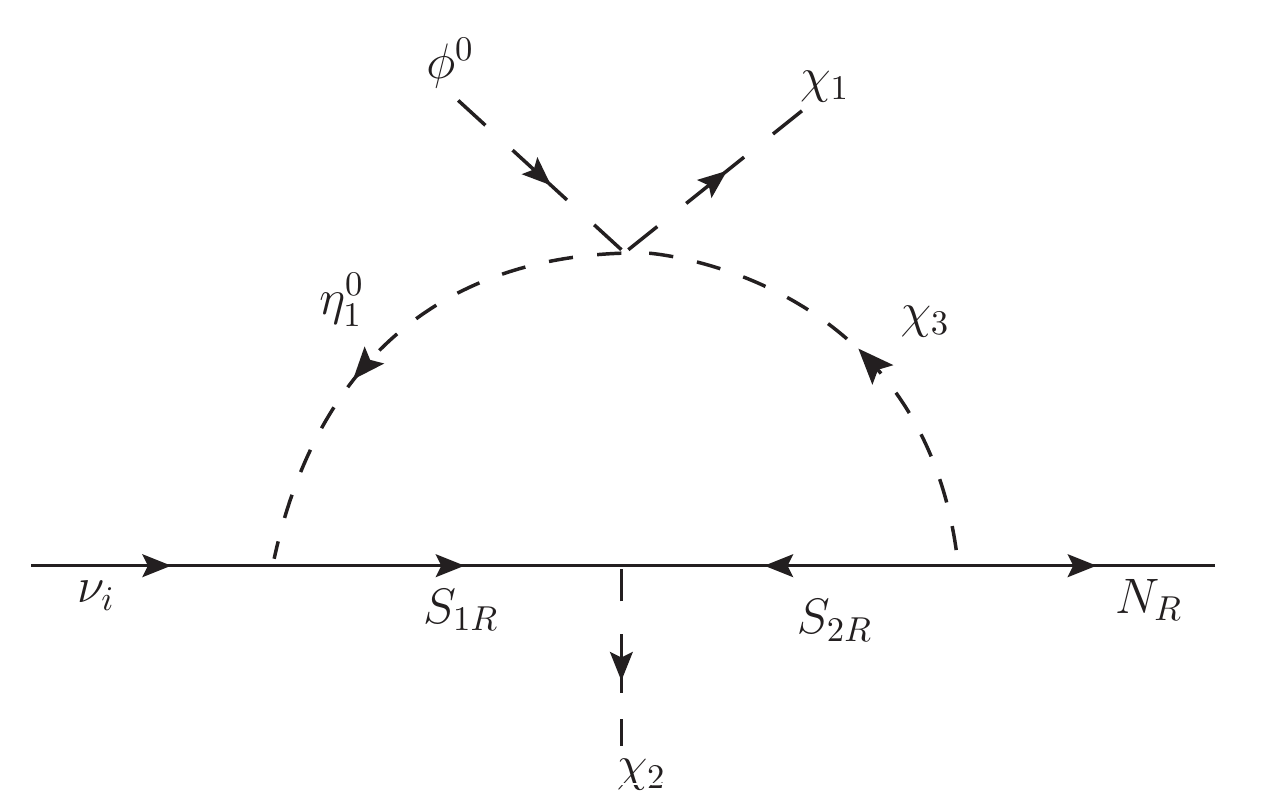}
 \caption{One-loop active-sterile neutrino mixing \cite{Adhikari:2015woo}.}
 \label{s1s2mixing}
 \end{figure}

There are eight real scalar fields, spanning $\sqrt{2}Re(\eta_{1,2}^{0})$, $ \sqrt{2}Im(\eta_{1,2}^{0})$, $\sqrt{2}Re(\chi_{3}^{0})$, $\sqrt{2}Im(\chi_{3}^{0})$, $\sqrt{2}Re(\xi^{0})$, $\sqrt{2}Im(\xi_{0})$ with mass eigenstate as $\zeta_{l}$ with mass $m_{l}$. These fields are present in one loop diagram giving radiative masses to active and sterile neutrinos. For details about the one loop diagrams giving masses to active and sterile neutrinos, we refer readers to Ref. \cite{Adhikari:2015woo}. However, we have shown the one loop diagram in Fig. \ref{s1s2mixing} gives rise to active and sterile neutrino mixing. 

Apart from three light active neutrinos, $N_R$ plays the role of fourth neutrino as sterile in this model as mentioned earlier. The masses of active and sterile neutrinos are given as
\begin{align}
{({\cal M}_\nu)^{(2)}_{ij}} = { h_{i1}^{\eta_{2}}  h_{j1}^{\eta_{2}} \over 16 \pi^2} 
\sum_{k} (z_{3k})^2 m_{\psi_k} \;A_{1}
+ { h_{i2}^{\eta_{2}}  h_{j2}^{\eta_{2}} \over 16 \pi^2} \sum_k (z_{4k})^2 m_{\psi_k}\;A_{2}
\label{activea1}
\end{align}
where $A_{1}=\sum_l [(y^R_{2l})^2 F(x_{lk}) - 
(y^I_{2l})^2 F(x_{lk})] $ and $A_{2}=\sum_l [(y^R_{2l})^2 F(x_{lk}) - (y^I_{2l})^2 F(x_{lk})]$\\
$\Sigma^0_{1R} = \sum_k z_{3k} \psi_k$, 
$\Sigma^0_{2R} = \sum_k z_{4k} \psi_k$, 
$\sqrt{2} Re(\eta_2^0) = \sum_l y^R_{2l} \zeta_l$, 
$\sqrt{2} Im(\eta_2^0) = \sum_l y^I_{2l} \zeta_l$, 
with $\sum_k (z_{3k})^2 = \sum_k (z_{4k})^2 = \sum_l 
(y^R_{2l})^2 = \sum_l (y^I_{2l})^2 = 1$, and $x_{lk} = m_l^2/m_{\psi_k}^2$ and $F(x_{lk})=x_{lk}\ln{x_{lk}}/(x_{lk}-1)$. Equation (\ref{activea1}) is the contribution to the active neutrino masses from $\Sigma_{1R}$ and $\Sigma_{2R}$. 

Let  $\bar{S}_{1R} \nu_i \eta_1^0$ coupling be $ h_{i1}^{\eta_{1}}$, 
then the contribution to ${\cal M}_\nu$ is given by
\begin{equation}
({\cal M}_\nu)_{ij} = { h_{i1}^{\eta_{1}}  h_{j1}^{\eta_{1}} \over 16 \pi^2} \sum_k 
(z_{1k})^2 m_{\psi_k}\;A
\label{activemass}
\end{equation}
where $A=\sum_l [(y^R_{1l})^2 F(x_{lk}) - (y^I_{1l})^2 F(x_{lk})],$ and
$S_{1R} = \sum_k z_{1k} \psi_k$,  
$\sqrt{2} Re(\eta_1^0) = \sum_l y^R_{1l} \zeta_l$, 
$\sqrt{2} Im(\eta_1^0) = \sum_l y^I_{1l} \zeta_l$, with 
$\sum_k (z_{1k})^2 = \sum_l (y^R_{1l})^2 = \sum_l (y^I_{1l})^2 = 1$. We have assumed that the first contribution to active neutrino mass shown in Eq.
(\ref{activea1}) is somewhat lesser than the second contribution shown in Eq. (\ref{activemass}) which gives masses to heavier neutrinos and the combination gives rise to appropriate mixing among different active neutrinos. We have considered only the mass scale for active neutrinos in our work and will be concerned with only Eq. (\ref{activemass}).

Let  $S_{2R} N_R \chi_3^0$ coupling be $h_{23}^{N}$, then the mass of sterile neutrino is given as
\begin{equation}
m_{NN} = { h_{23}^{N}  h_{23}^{N} \over 16 \pi^2} \sum_k 
(z_{2k})^2 m_{\psi_k} \;B
\label{sterilemass}
\end{equation}
where $B=\sum_l [(y^R_{3l})^2 F(x_{lk}) - (y^I_{3l})^2 F(x_{lk})]$ and $S_{2R} = \sum_k z_{2k} \psi_k$,  
$\sqrt{2} Re(\chi_3^0) = \sum_l y^R_{3l} \zeta_l$, 
$\sqrt{2} Im(\chi_3^0) = \sum_l y^I_{3l} \zeta_l$, with 
$\sum_k (z_{2k})^2 = \sum_l (y^R_{3l})^2 = \sum_l (y^I_{3l})^2 = 1$. \\
 
 The active-sterile neutrino mixing is possible because of interaction $h_{12}^{S}S_{1R}S_{2R}\bar{\chi_2^0}$ as shown in Fig. 1 and non-diagonal mass matrix elements related to mixing in active sterile neutrino mass matrix is given as 
\begin{equation}
m_{\nu N} = {h_{i1}^{\eta_{1}} (h_{23}^{N}) \over 16 \pi^2} \sum_k z_{1k} 
z_{2k}\;m_{\psi_k}\;C 
\label{mixingmass}
\end{equation}
where $ C=\sum_l [y_{1L}^R y_{3l}^R F(x_{lk}) - y^I_{1l} y^I_{3l} F(x_{lk}]$
and $\sum_k z_{1k} z_{2k} = \sum _l y^R_{1l} y^R_{3l} = \sum_l y^I_{1l} 
y^I_{3l} = 0$.\\
$A,B$ and $C$ are loop factors corresponding to the one loop diagrams that gives masses and mixing of neutrinos. Based on recent global fit \cite{gfit} of 
neutrino oscillation experiment with sterile neutrino in 3+1 scheme, the best fit values are: $\Delta m_{41}^2=1.3 \; \mbox{eV}^2, \; |U_{e4}|=0.1$
and $|U_{\mu 4}|\lesssim 10^{-2} $. Also taking into account the cosmological constraint on sum of three active neutrino masses \cite{Vagnozzi:2018jhn,Choudhury:2018byy}  we consider active neutrino masses, sterile neutrino mass and active-sterile mixing as
\begin{equation}
   \left({\cal M}_{\nu } \right)_{ij}\sim 0.1\mbox{eV},\;\; \;
   {\cal M}_{NN} \sim 1.14\mbox{eV},\;\;\;{\cal M}_{\nu N} \sim 0.114 \mbox{eV}
   \label{bestfitvalue}
\end{equation}
  The product of the mixing matrix element and $m_{\psi_k}$ which are present in Eqs. (\ref{activemass}), (\ref{sterilemass}) and  (\ref{mixingmass}) can be rewritten in terms of the mixing angle $\theta$ and the mass gap parameter  $\Delta = (m_{\psi_{2}}-m_{\psi_{1}})/m_{\psi_{1}}$ after we consider
 $z_{11}=\cos\theta$,  $z_{12}=-\sin\theta$,  $z_{21}=\sin\theta$ and  $z_{22}=\cos\theta$. Following these we can write 
\begin{align}
\sum_{k}z_{1k} \; z_{2k}\; m_{\psi_{k}}= \frac{ \Delta\; m_{\psi_{1}}\;\sin2\theta}{2} \nonumber \\
\sum_{k}\left( z_{2k}\right)^{2}\; m_{\psi_{k}}= m_{\psi_{1}}\left(1+ \Delta \;\cos^{2}\theta\right) \nonumber \\
\sum_{k}\left( z_{1k}\right)^{2}\; m_{\psi_{k}}= m_{\psi_{1}}\left(1+ \Delta \;\sin^{2}\theta\right).
\label{zangle}
\end{align}
 Using  eqs. \eqref{bestfitvalue} and \eqref{zangle} for active neutrino mass scale and the mixing of sterile neutrino $U_{e4}$  we can write equation (\ref{activemass}), (\ref{sterilemass}) and (\ref{mixingmass}) in terms of $\Delta$ and $\theta$ parameters and can be written as
\begin{equation}
 { h_{11}^{\eta_{1}}  h_{11}^{\eta_{1}} \over 16 \pi^2}\; m_{\psi_{1}}\left(1+ \Delta \;\sin^{2}\theta\right)\;A \approx 0.1
\label{activemass2}
\end{equation}
\begin{equation}
 { h_{23}^{N}  h_{23}^{N} \over 16 \pi^2}\; m_{\psi_{1}}\left(1+ \Delta \;\cos^{2}\theta\right)\;B \approx 1.14
\label{sterilemass2}
\end{equation}
\begin{equation}
 { h_{11}^{\eta_{1}}  h_{23}^N\over 16 \pi^2}\; m_{\psi_{1}}\left(\frac{(\Delta \;\sin{2\theta}}{2}\right)\;C \approx 0.1
\label{mixingmass2}
\end{equation}
 \begin{figure}[h]
\centering
\includegraphics[scale=0.8]{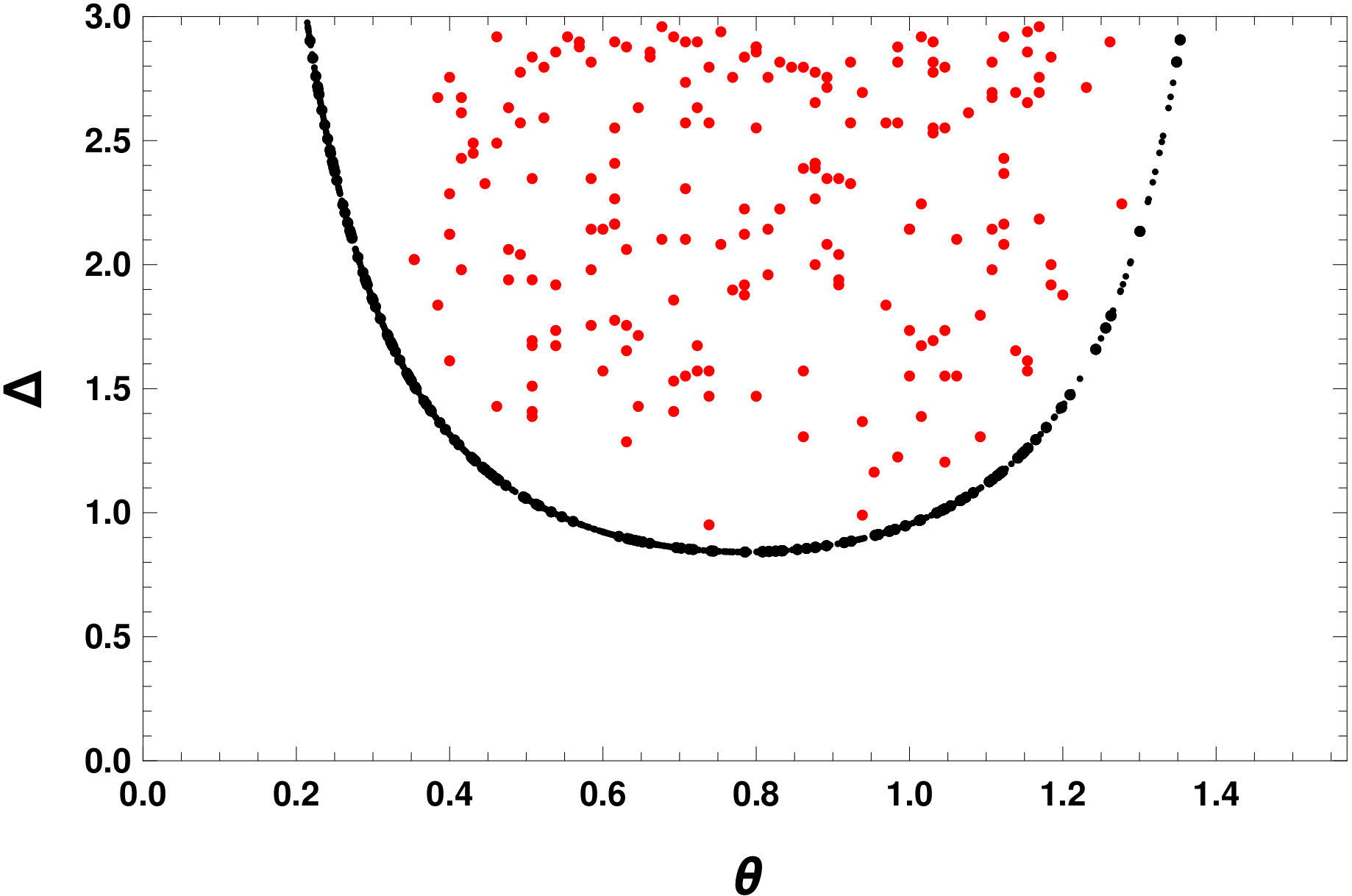}
\caption{Allowed region of $\Delta$ and $\theta$ (in radian) from active-sterile neutrino masses and mixing.}
\label{figthetadel2}
\end{figure}
 If we consider $ \sum z^2 m_{\psi_k}$ of equation \eqref{zangle} of the order of 1 TeV and couplings $h \sim 0.1$ then from  equations  (\ref{activemass2}), (\ref{sterilemass2}) and  (\ref{mixingmass2}), it is found that $A$, $B$ and $C$ are in the range of $10^{-8}$ to $10^{-9}$. However, considering the variation of two couplings $ h_{11}^{\eta_{1}}$ and $h_{23}^N$ in the range of 0.05 to 1  and   $A$, $B$ and $C$ in the range of $10^{-9}$ to $10^{-6}$  in  equations  (\ref{activemass2}), (\ref{sterilemass2}) and  (\ref{mixingmass2}) we obtain 
 allowed region of $\theta $ and $\Delta$ as shown in the Fig. \ref{figthetadel2}.  In Fig. \ref{figthetadel2} we have considered two different conditions among $A$, $B$ and $C$ : 1) $A=B=C$ and 2)
 $C$ $<$ $A$, $B$. The condition 1) gives the allowed almost semicircle outer line whereas condition 2) gives region inside covered by that almost semicircle line. One may note here that $A$ and $B$ as mentioned just after equation (10) and (11) are very similar in nature with sum over the product of the mixing matrix elements of $y$ are 1 for both A and B, whereas for $C$ as mentioned just after equation (12) due to orthogonality condition the sum over the product of the mixing matrix elements of $y$ vanishes. Due to this difference, $C$ is expected to be lesser than both $A$ and $B$. In that sense we should consider the proper allowed region of $\Delta$ and $\theta$ as that given by the region inside the almost semicircle line in Fig. \ref{figthetadel2} and $\Delta \geq 1$ is found to be preferred.

For active and sterile neutrino mixing,  one important conclusion follows from both Fig. 1 and  \ref{figthetadel2}  is that there is necessarily  non-zero mixing $\theta$ between $S_{1R}$ and $S_{2R}$ as otherwise there will be zero contribution from Fig. 1. Fig. \ref{figthetadel2} however, shows apart from non-zero mixing, the simultaneous constraint on both $\theta$ and $\Delta$.  These imply that apart from annihilation of two dark matter fields ($\psi_1$) into
 SM fermion and antifermion pair, there is co-annihilation of dark matter field $\psi_1$ with other next heavier Majorana field $\psi_2$ through the interactions mentioned in equation \eqref{12}.
 
\section{Dark Matter ($ \psi_1 $) relic density }
Relic density is obtained from the Boltzmann equation \cite{Beltran:2008xg} governing the evolution of number density of the DM with the  thermally averaged cross section  for the process $\psi_{1}\psi_{1}\rightarrow f\bar{f}$. The Boltzmann equation is written as:
\begin{equation}
\dot{n_{\psi_1}}+3Hn_{\psi_1}=<\sigma v>((n_{\psi_1}^{eqb})^2-n_{\psi_1}^2)
\end{equation}
where $n_{\psi_1}$ is the number density  and $n_{\psi_1}^{eqb}$ is thermal equilibrium number density of the DM particle. $H$ is Hubble expansion rate of the universe and $\langle\sigma v\rangle$ is the thermally averaged cross section for the process $\psi_{1}\psi_{1}\rightarrow f\bar{f}$ and is given by \cite{Gondolo:1990dk}
\begin{equation}
<\sigma v>=\frac{1}{8m_{\psi_1}^4T K_2^2(m/T)}\int_{4m_{\psi_1}^2}^{\infty} \sigma(s-4m_{\psi_1}^2) \sqrt{s}K_{1}(\sqrt{s}/T)ds
\end{equation}
where $K_{1} , K_{2}$ are  modified Bessel functions of first and second kind respectively. Here $s$ is the centre of mass energy squared. The thermally averaged cross section can be expanded in powers of relative velocity of two dark matter particle to be scattered and is written as $ <\sigma v> = a +b  v^2$. Numerical solution of the above  Boltzmann equation gives \cite{Kolb:1990vq}
\begin{equation}
\Omega_{\psi_1} h^2\approx\frac{1.04\times10^9 x_{f}}{M_{pl}\sqrt{g_{*}}(a+3b/x_{f})}
\label{omegaold}
\end{equation}
where $x_{f}=m_{\psi_1}/T_{f}$ , $T_{f}$  is the freeze-out temperature, $g_{*}$ is the number of relativistic degrees of freedom at the time of freeze out. $x_{f}$ can be find out from 
\begin{equation}
x_{f}=\ln\frac{0.038 \; M_{Pl}\; m_{\psi_1}<\sigma v>}{g_{*}^{1/2}x_{f}^{1/2}}.
\label{xfold}
\end{equation}
 
 We assume $\psi_1$ to be lighter than $\psi_2$ and it plays the role of dark matter.  The lighter $\psi_{{1}}$ will be pair annihilated to SM fermions and anti-fermions through the extra gauge boson mediator but not $Z$ boson as we are considering $Z-X$ mixing to be zero. This pair annihilation has been considered above. However, it could also co-annihilate with  $\psi_{{2}}$. 
Both  annihilation and  co-annihilation  cross sections  could control the relic abundance of the dark matter $\psi_1$ \cite{Griest,Cannoni:2015wba}. To take into account co-annihilation we discuss the necessary modifications in the Boltzmann equation below.

 If the mass difference between $\psi_{{1}}$ and $\psi_{{2}}$ is very large, $\psi_{{2}}$ will be out of thermal equilibrium much earlier than $\psi_{{1}}$ and the co-annihilation will not play significant role in the evolution of the number density of $\psi_{{1}}$. However, we consider the case where the mass difference may not to be too large. In that case, we consider the annihilation as well as co-annihilation channel in the coupled Boltzmann equation to find out the evolution of the number density of $\psi_{{1}}$  and hence  find the relic density of dark matter.  Using the formalism of Ref.\cite{Griest} we can reduce the system of $2$ Boltzmann equations governing number densities $n_1$ and $n_2$ of $\psi_{{1}}$ and $\psi_{{2}}$ respectively into one Boltzmann equation which governs the evolution of $n=n_1+n_2$ in the early universe as given below:  
 \begin{equation}
\dot{n}=-3Hn-\sum_{i,j=1}^{2}<\sigma_{ij} v>((n_{i}n_{j}-n_{i}^{eq}n_{j}^{eq})
\end{equation}
 where $<\sigma_{ij} v>$ is the thermally averaged scattering cross section  for the process $\psi_{i}\psi_{j}\rightarrow f \bar{f}$. This equation can be further simplified as 
 
 \begin{equation}
\dot{n}=-3Hn-\sum_{i,j=1}^{2}<\sigma_{eff} v>((n^{2}-(n^{eq})^{2})
\end{equation}
where $\sigma_{eff}$ is given as
\begin{equation}
    \sigma_{eff}\approx\sum_{i,j=1}^{2}\sigma_{ij}\frac{g_{i}g_{j}}{g_{eff}^2}(1+\Delta_{i})^{3/2}(1+\Delta_{j})^{3/2}\exp^{-x(\Delta_{i}+\Delta_{j})}.
    \label{sigeff}
\end{equation}
Here   $x=m_{\psi_1}/T$ and $\Delta_{i}=\frac{m_{\psi_{i}}-m_{\psi_{1}}}{m_{\psi_{1}}}$ . Then $\Delta_1=0$ by definition. Later on, $\Delta_2 $ is written  as $\Delta$. $g_{i}$ is the internal degrees of freedom of the interacting particles and  $g_{eff}$ is defined as 
\begin{equation}
    g_{eff}=\sum_{i=1}^{2}g_{i}(1+\Delta_{i}){3/2}\exp^{-x\Delta_{i}}
\end{equation}

For comparison with the general WIMP formulas, we have Taylor expanded the thermally averaged cross-sections:
\begin{align}
<\sigma_{ij} v>=a_{ij}+b_{ij}v^{2} \,\,\, ; \,\,\, <\sigma_{eff} v>=a_{eff}+b_{eff}v^{2}
\label{sigma}
\end{align}
where $a_{eff}$ and $b_{eff}$ are given by
 \begin{equation}
    a_{eff}\approx\sum_{i,j=1}^{2}a_{ij}\frac{g_{i}g_{j}}{g_{eff}^2}(1+\Delta_{i})^{3/2}(1+\Delta_{j})^{3/2}\exp^{-x(\Delta_{i}+\Delta_{j})}.
    \label{aeff}
\end{equation}
\begin{equation}
    b_{eff}\approx\sum_{i,j=1}^{2}b_{ij}\frac{g_{i}g_{j}}{g_{eff}^2}(1+\Delta_{i})^{3/2}(1+\Delta_{j})^{3/2}\exp^{-x(\Delta_{i}+\Delta_{j})}.
    \label{beff}
\end{equation}

 The phase space integration part  for all the process of  $\psi_{i}\psi_{j}\rightarrow f\bar{f}$ are almost same for cross section calculation and the difference in the cross sections are mainly due to the strength of couplings $g_{ij}$ in which both the indices $i,j$ runs from 1 to 2. Because of this we can write 
\begin{equation}
    \frac{\sigma_{11}}{|g_{11}|^{2}}\approx \frac{\sigma_{12}}{|g_{12}|^{2}}\approx \frac{\sigma_{21}}{|g_{21}|^{2}}\approx \frac{\sigma_{22}}{|g_{22}|^{2}}
\end{equation}
Using this approximation $\sigma_{eff}$ in equation (\ref{sigeff}) can be written in terms of $\sigma_{11}$ as
\begin{equation}
    \sigma_{eff}=\frac{g_{i}g_{j}}{g_{eff}^{2}}\left[1+2 \; r_{12}\frac{g_{12}^{2}}{g_{11}^{2}}+r_{22}\frac{g_{22}^{2}}{g_{11}^{2}} \right]\sigma_{11}
\label{sigeff2}
\end{equation}
Here $r_{12}=(1+\Delta)^{3/2}e^{-x\Delta}\; , r_{22}=(1+\Delta)^{3}e^{-2x\Delta}\;\mbox{and}\;\Delta=(m_{\psi_{2}}-m_{\psi_{1}})/m_{\psi_{1}}$ and $\sigma_{11}$ is annihilation cross section of $\psi_1\psi_1\rightarrow f\bar{f}$ and $<\sigma_{11} v>$ can be Taylor expanded in the form of $a_{11}+b_{11}v^{2}$. For Majorana fermions $g_{i}=g_{j}=2\;\mbox{(internal degrees of freedom)}$ and from equation \eqref{12} the couplings involved in these annihilation and co-annihilation channels are;
\begin{align}
    g_{11}&=g_{S_{1R}a}\cos^2\theta +g_{S_{2R}a}\sin^2\theta\nonumber\\
    g_{22}&=g_{S_{2R}a}\cos^2\theta +g_{S_{1R}a}\sin^2\theta\nonumber\\
    g_{12}&=g_{21}=\frac{1}{2}\sin2\theta(g_{S_{1R}a}-g_{S_{2R}a}) 
    \label{coupling}
\end{align}
In presence of co-annihilation of dark matter $\psi_1$ with $\psi_2$ equation \eqref{omegaold} and \eqref{xfold} will be modified as;
\begin{equation}
\Omega_{\psi_1} h^2\approx\frac{1.04\times10^9 x_{f}}{M_{pl}\sqrt{g_{*}}(a_{11}I_{a}+3b_{11}I_{b}/x_{f})}
\label{omegaf}
\end{equation}
where
\begin{equation}
I_{a}=\frac{x_{f}}{a_{11}}\int_{x_{f}}^{\infty}x^{-2}a_{eff}dx
   \quad\text{and}\quad 
I_{b}=\frac{2x_{f}^{2}}{b_{11}}\int_{x_{f}}^{\infty}x^{-3}b_{eff}dx
\end{equation}
and $x_f$ can be obtained from \begin{equation}
x_{f}=\ln\frac{0.038M_{Pl}m_{\psi_{1}}<\sigma_{eff} v>}{g_{*}^{1/2}x_{f}^{1/2}}.
\end{equation}

 Following  \cite{Berlin:2014tja} the annihilation cross section of Majorana fermion to SM $f\bar{f}$ through s-channel  mediated by X boson is given as ,
\begin{align}
&\sigma_{11} = \frac{n_{c}}{12 \pi s \left[\left(s-M_{X}^2\right)^2+M_{X}^{2}\Gamma_{X}^{2}\right]}\sum_{f}\sqrt{\frac{1-4m_{f}^{2}/s}{1-4m_{\psi_{1}}^{2}/s}} \bigg[ g_{fa}^{2}g_{11}^{2}\left(4m_{\psi_{1}}^{2}\left[m_{f}^{2}\bigg(7-\frac{6s}{M_{X}^{2}}+\frac{3s^2}{M_{X}^{4}}\right)-s\right] \nonumber \\&+s(s-4m_f^{2})\bigg)+g_{fv}^{2}g_{11}^{2}(s+2m_{f}^{2})(s-4m_{\psi_{1}}^{2}) \bigg]
\label{23}
\end{align}
where $n_{c}= 3$ when $f$ stands for quarks and $n_{c}= 1$ when $f$ stands for leptons and $s\approx 4m_{\psi_1}^{2}+ m_{\psi_1}^{2}v^{2} $. In above equation
$\Gamma_{X}=\Gamma_{X_1}+\Gamma_{X_2}$, is the total decay width of extra gauge boson $X$ where $\Gamma_{X_1}$ is the decay width of extra gauge boson $X$ to SM fermion anti-fermion pair and $\Gamma_{X_{2}}$  is the decay width of extra gauge boson decaying to $\psi_{i}\psi_{j}$ Majorana fermions.
\begin{eqnarray}
&&\Gamma_{X_1}\equiv\sum_{f}\Gamma (X\rightarrow f\bar{f})=\sum_{f}\frac{n_{c}M_{X}}{12\pi S}\sqrt{1-\frac{4m_{f}^2}{M_{X}^2}}\left[g_{fa}^2\left(1-\frac{4m_{f}^2}{M_{X}^2}\right)+g_{fv}^2\left(1-2\frac{m_{f}^2}{M_{X}^2}\right)\right]\nonumber\\
&&\Gamma_{X_2}=\sum_{i,j}\Gamma (X\rightarrow \psi_{i}\bar{\psi_{j}})=\sum_{i,j}\frac{M_{X}g_{ij}^2}{12\pi S}\left[1-\frac{4m_{\psi_i}m_{\psi_j}}{M_{X}^2}\right]^{3/2}
\end{eqnarray}
where S = 1 (2) for (in)distinguishable final state particles.
In the notation of equation \eqref{sigma}, $<\sigma_{11} v> =a_{11}+b_{11}v^2$ , where 
\begin{eqnarray}
&& a_{11}=\frac{n_{c} \, g_{fa}^{2} \, m_{f}^{2} \, g_{11}^2 \, m_{\psi_{1}}}{24\pi\left[(M_{X}^{2}-4m_{\psi_{1}}^{2})^{2}+M_{X}^{2}\Gamma_{X}^{2}\right]}\sqrt{1-\frac{m_{f}^{2}}{m_{\psi_{1}}^{2}}}\left(12-96\frac{m_{\psi_{1}}^{2}}{M_{X}^{2}}+192\frac{m_{\psi_{1}}^{4}}{M_{X}^{4}}\right),  \nonumber \\
&& b_{11}=a_{11}\left[-\frac{1}{4}+\frac{m_{f}^{2}}{8(m_{\psi_{1}}^{2}-m_{f}^{2})}-\frac{M_{X}^4-16M_{X}^2m_{\psi_{1}}^2+48m_{\psi_{1}}^4}{4((M_{X}^2-4m_{\psi_{1}}^2)^2+M_{X}^2\Gamma_{X}^2)}\right.  \nonumber\\
 &&\left. +\frac{\left(-4+2\frac{g_{fv}^{2}}{g_{fa}^2}+4\frac{m_{\psi_{1}}^2}{m_{f}^2}+4\frac{g_{fv}^{2}m_{\psi_{1}}^2}{g_{fa}^2m_{f}^2}-24\frac{m_{\psi_{1}}^2}{M_{X}^2}+96\frac{m_{\psi_{1}}^4}{M_{X}^4}\right)}{\left(12-96\frac{m_{\psi_{1}}^2}{M_{X}^2}+192\frac{m_{\psi_{1}}^4}{M_{X}^4}\right)}\right].
\end{eqnarray}
Mass $M_{X}$ of the extra gauge boson in above expressions is given by equation \eqref{massxboson} and $g_{11}$ corresponds to axial-vector coupling as follows from equations \eqref{coupling} and \eqref{12}.

 \section{Allowed region of dark matter mass, $X$ boson mass and its gauge coupling} 
 Large  Hadron Collider (LHC) operating at $\sqrt{s}=13 \; \mbox{TeV}$, have searched for new phenomena \cite{Aaboud:2017yvp,Aaboud:2017buh,cms} which are resonant as well as non-resonant and in which the final state is dilepton/dijet. This is a robust test for  all theories  beyond the SM. ATLAS at LHC has obtained allowed region of coupling $g_X$ of quarks and leptons with extra gauge boson mass $M_X$ in Fig. (4) of \cite{Aaboud:2017yvp} and Fig. (5a) of  \cite{Aaboud:2017buh} at $95\%$ confidence level. However, these two Fig.s in reference \cite{Aaboud:2017yvp} and \cite{Aaboud:2017buh} do not differ too much. We have considered the allowed region of $g_X$ and $M_X$ in Fig. (5a) of ref \cite{Aaboud:2017buh} in our numerical analysis. $g_X$  coupling in our paper is related to  coupling $\gamma^{'}$ of this Fig. as $g_X\approx0.463 \gamma^{'} $.

For our numerical analysis $U(1)$ charges are fixed by considering  $n_1=n_4= 1/\sqrt{2}$ (normalizing $n_1^2+n_4^2=1$) which satisfies zero mixing condition of extra gauge boson $X$ with SM $Z$ boson. We have considered two values of $\Delta \; \mbox{as}  \; 1 \; \mbox{and} \;2$ and also different values of $\theta \; \mbox{as} \; 0.4 \;, \;\pi/4 \; \mbox{and}~ 1.2$ as allowed by Fig. \ref{figthetadel2}. 

\begin{figure}[!h]
\begin{center}
\begin{tabular}{c c}
 \includegraphics[width=7.5cm,height=5.5cm]{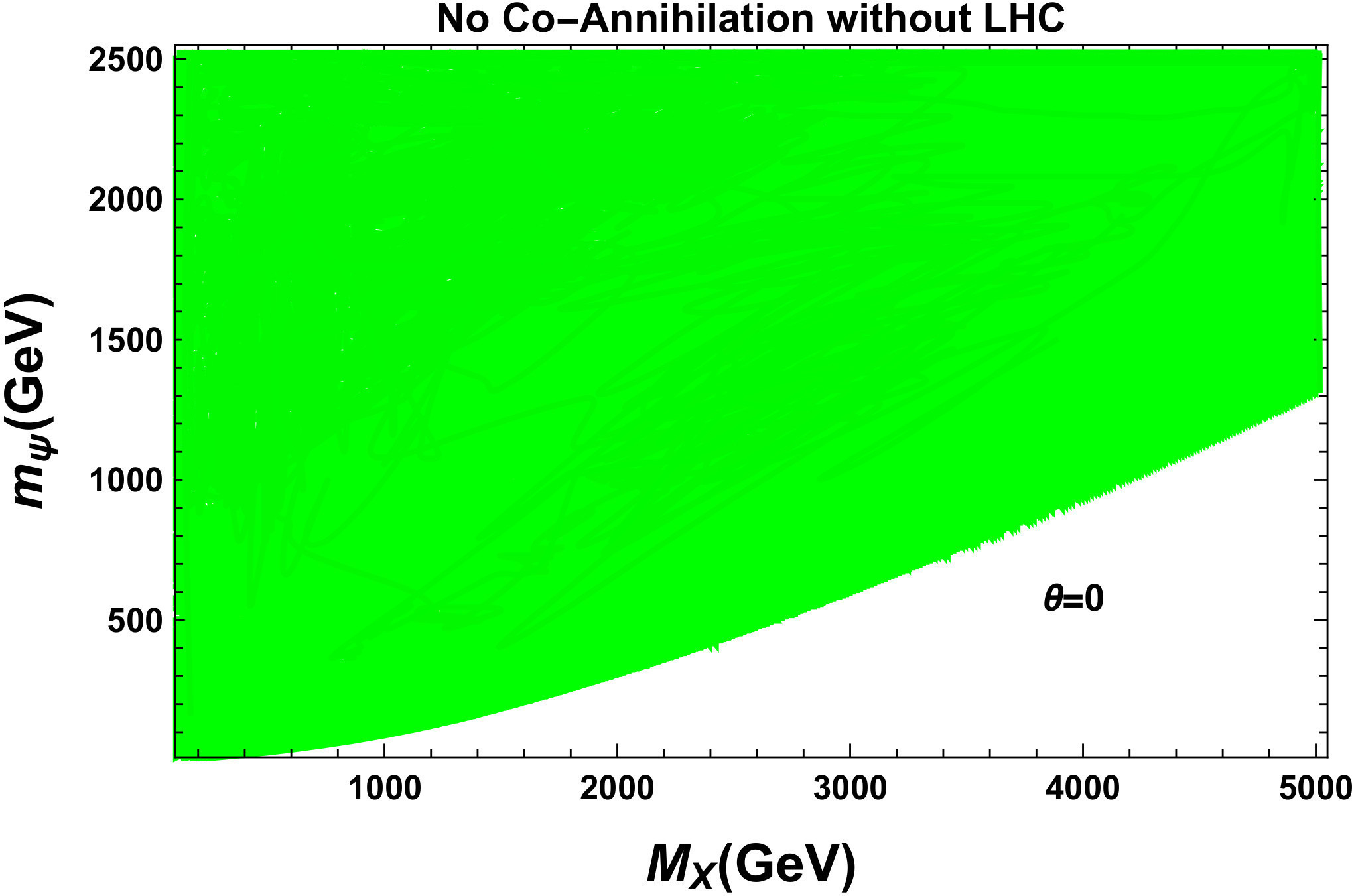}&
 \includegraphics[width=7.5cm,height=5.5cm]{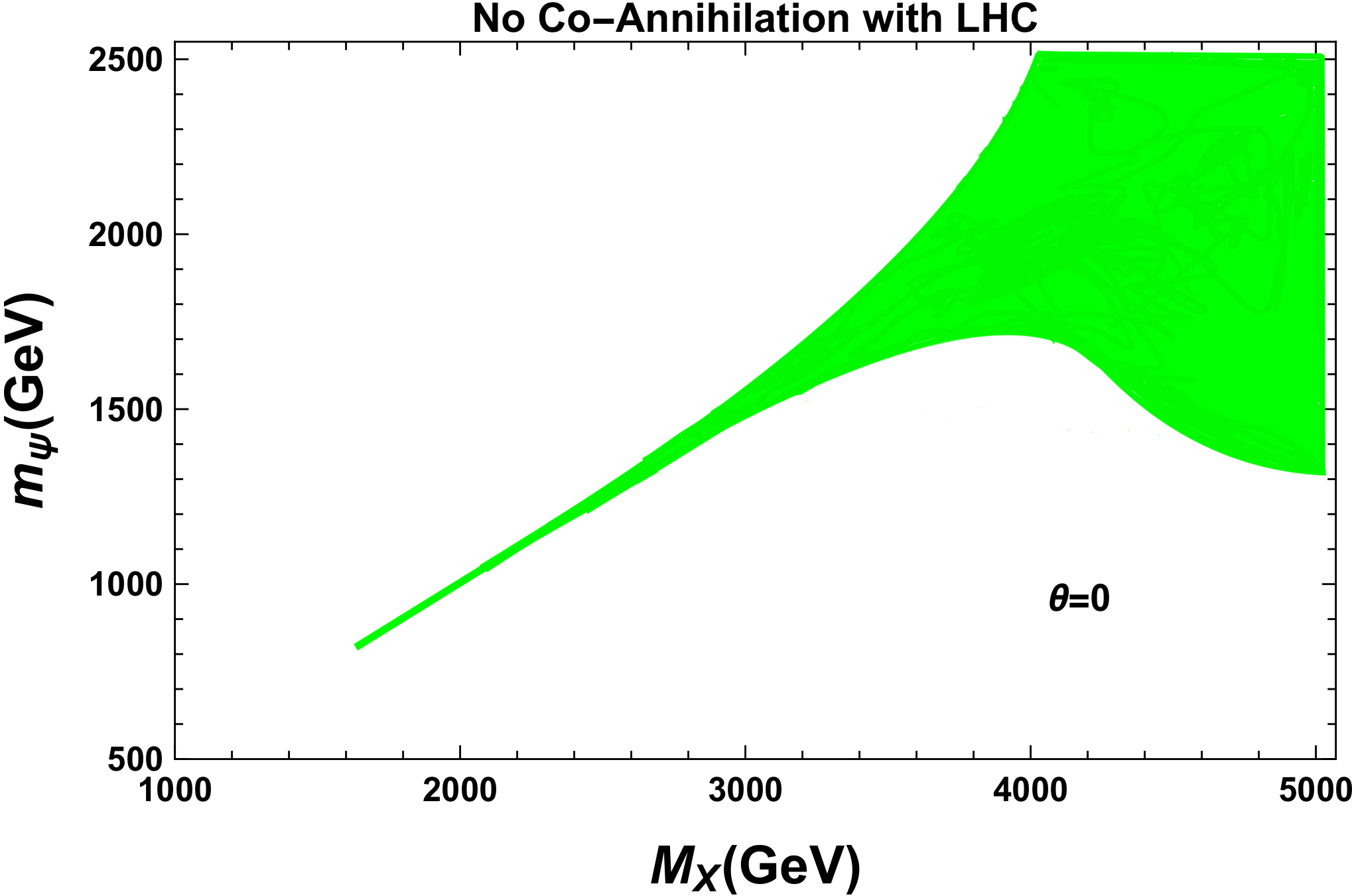}\\ 
(a) & (b) \\
\end{tabular}
\end{center}
\caption{Allowed green region in the $M_{X}$ and $m_{\psi}$ plane for no co-annihilation channel satisfies  PLANCK relic density bound $\Omega h^{2}\in(0.1188,0.1212)$. The Fig. (a) and Fig. (b) corresponds to no LHC constraints and LHC constraints on $M_{X}$ and $g_{X}$ respectively.}
\label{noco}
\end{figure}

In later part we have used the symbol $m_\psi$ for the dark matter mass $m_{\psi_1}$. In Fig. \ref{noco} we first show the allowed region in $M_X - m_{\psi}$ plane for only annihilation channel ($\psi_{1}\psi_{1} \rightarrow f\bar{f}$) but no co-annihilation of dark matter fermion. As discussed earlier this will not in general correspond to active and sterile neutrino mixing in the model considered
by us because of no co-annihilation channel. For comparing the allowed region with and  without LHC constraint as shown in Fig. \ref{noco} we have considered the variation of $g_{X}$ over the range (0.005-0.7), $m_{\psi}$ up to 2.5 TeV and $M_{X}$ up to 5 TeV, as this is the range considered by LHC. The other two parameters $\theta$ and $\Delta$ are zero as this Fig. is only for annihilation case.  In plotting the Fig. \ref{noco}(a) we have considered only the constraint coming from relic abundance on dark matter from PLANCK  2018 $\Omega h^2\in(0.1188,0.1212)$\cite{Aghanim:2018eyx}. But in Fig. \ref{noco}(b) we have also considered the constraint 
on $g_X$ and $M_X$ given by ATLAS collaboration \cite{Aaboud:2017yvp,Aaboud:2017buh} at LHC. Comparing Fig. \ref{noco} (a) and (b) one can see  that LHC constraint significantly reduces the allowed region  of dark matter mass for lower values of $M_X < 4000$ GeV and for further lower $M_X$ the allowed range of dark matter mass $m_{\psi}$ is further constrained with respect to no LHC constraint in Fig. \ref{noco}(a). For $M_X$  lesser than about 1600 GeV  and $m_\psi$ lesser than about 700 GeV it is difficult to get any allowed region.

\begin{figure}[h]
\begin{center}
\begin{tabular}{c c}
\includegraphics[width=7cm,height=4.5cm]{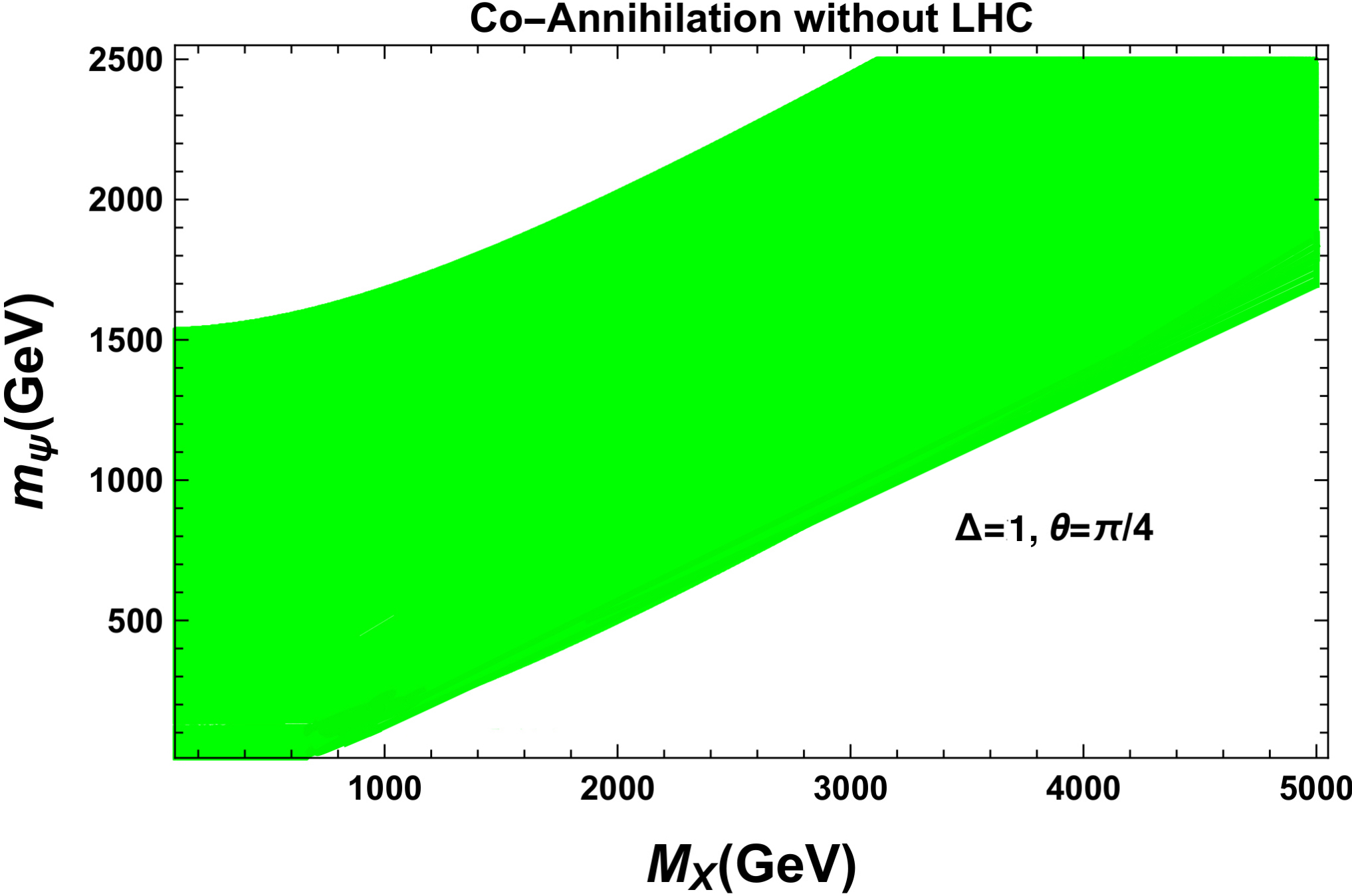}& 
\includegraphics[width=7cm,height=4.5cm]{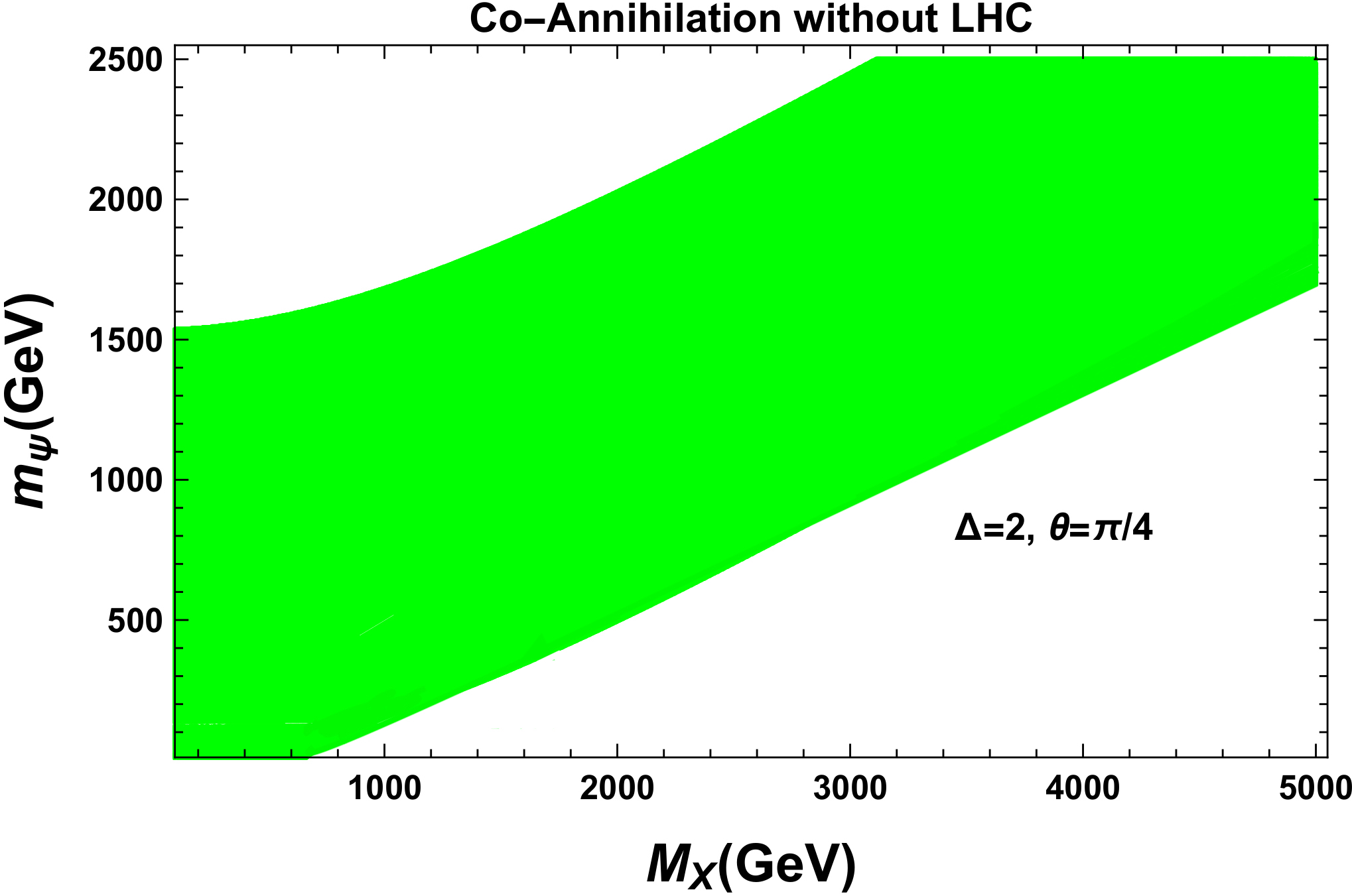}\\
(a) & (b)\\
\includegraphics[width=7cm,height=4.5cm]{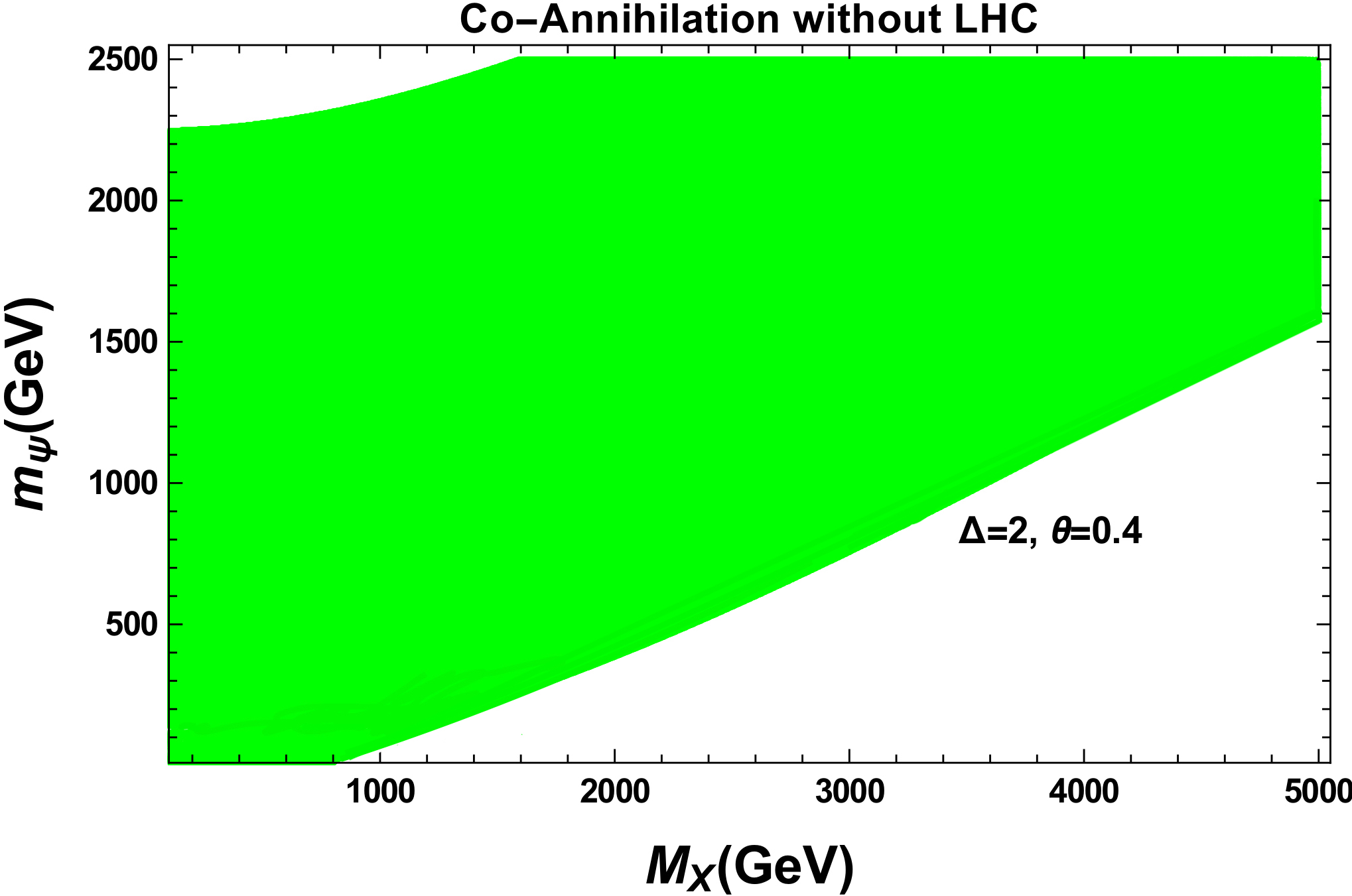}&
\includegraphics[width=7cm,height=4.5cm]{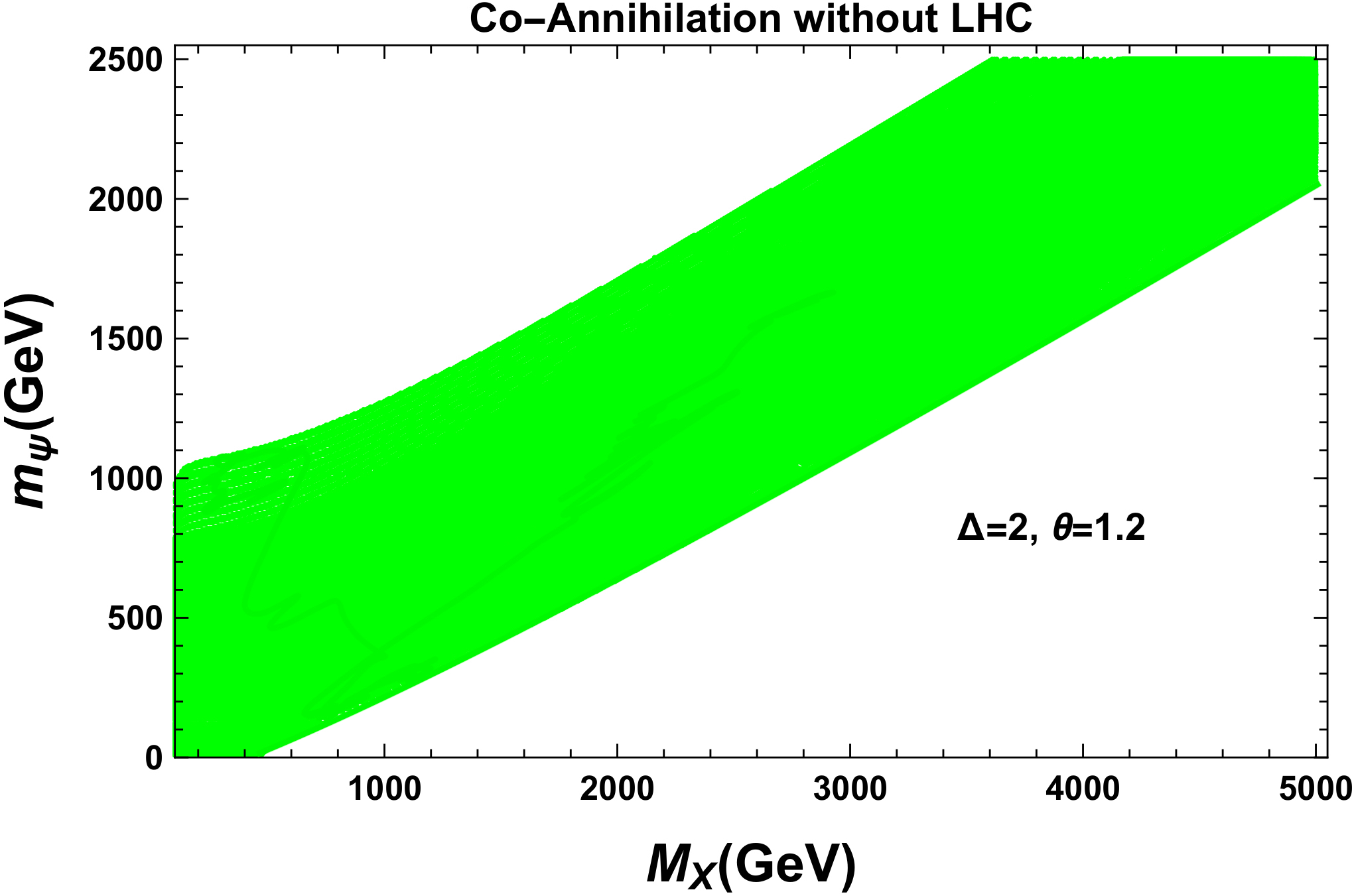}\\
(c) & (d)
\end{tabular}
\end{center}
\caption{Allowed green region in the $M_{X}$ and $m_{\psi}$ plane for co-annihilation channels with different choices of $\Delta =2$,  $\theta =0.4,~\pi/4$, 1.2  and $\Delta =1$, $\theta =\pi/4$ (in radian). The allowed region satisfies PLANCK relic density bound $\Omega h^{2}\in(0.1188,0.1212)$ .}
\label{cofig1}
\end{figure}

In Fig. \ref{cofig1} apart from considering constraint coming from relic abundance on dark matter from Planck 2018, we have also considered co-annihilation channel along with annihilation channel for the dark matter as required for satisfying active and sterile neutrino masses and mixing. As discussed in section III the preferred allowed region of
$\Delta$ and $\theta$ are shown in Fig. \ref{figthetadel2} inside the almost semicircular line. We consider in Fig. \ref{cofig1} different values of $\Delta$ and $\theta$ based on Fig. \ref{figthetadel2}. But no LHC constraint has been imposed.   

One can see after including co-annihilation channel for all cases in Fig. \ref{cofig1} there is significant constraint on allowed 
dark matter mass for any $M_X$ value in comparison to Fig. \ref{noco}(a) with no co-annihilation for which $\theta=0$. Significant change has come mainly due to non-zero $\theta$ value considered in Fig. \ref{cofig1} as the coupling $g_{ij}$ of $\psi_i \psi_j X$ as shown in equation \eqref{12} and (31) changes with the change in $\theta$ values. As for example, for $\theta=0$, $g_{12}$ 
is zero. The 2nd term (related to the process $\chi_1 \chi_2 \rightarrow f \bar{f} $) in $\sigma_{eff}$ in  equation \eqref{sigeff2} is zero for $\theta=0$ whereas this is non-zero for non-zero $\theta$ value. $\Delta$ plays the role of more suppression of the effect of this process on $\sigma_{eff}$ for its' higher values. The third term (related to the process $\chi_2 \chi_2 \rightarrow f \bar{f}$) in $\sigma_{eff}$ is more exponentially suppressed than the 2nd term for higher $\Delta$ values. So if $\Delta$ values are increased 
further than those considered in Fig. 4, there will be further lesser effect from 2nd and 3rd term in $\sigma_{eff}$.  For smaller $\Delta$ values much lesser than 1   there would have been more effect from the 2nd and 3rd term in $\sigma_{eff}$ in  equation \eqref{sigeff2} due to co-annihilation. One may note however, that active and sterile neutrino mixing constrains both $\Delta$ and $\theta$ simultaneously as shown in Fig. 2 and for $\Delta $ less than about 1, there is no allowed $\theta$ value
for co-annihilation to occur.  In Fig. \ref{cofig1} (a) and (b) we have chosen $\Delta=2,\; \theta=\pi/4$ and $\Delta=1,\; \theta=\pi/4$ respectively to see how much the allowed region in 
$M_X - m_{\psi}$ plane changes due to this variation of $\Delta$  for same $\theta  $  value. In fact, one can see that the change
is insignificant as both the Fig.s are almost same.  However, it is also expected that there will be  change in $\sigma_{eff}$ to some extent for variations in  $m_{\psi_1}$ values also as follows from \eqref{sigeff2} but this is 
subject to details of the Boltzman equations and the corresponding freeze-out temperature $T$ .  In Fig. \ref{cofig1} (c) and (d) we have changed $\theta$ values to 
$0.4$ and $1.2$ respectively with $\Delta$ fixed at 2.  Comparing Fig. \ref{cofig1} (b), (c) and (d), it is found that the allowed region changes significantly with such  variations of $\theta $ value. 

\begin{figure}[h!]
\centering
\begin{tabular}{c c}
\includegraphics[width=7cm,height=5cm]{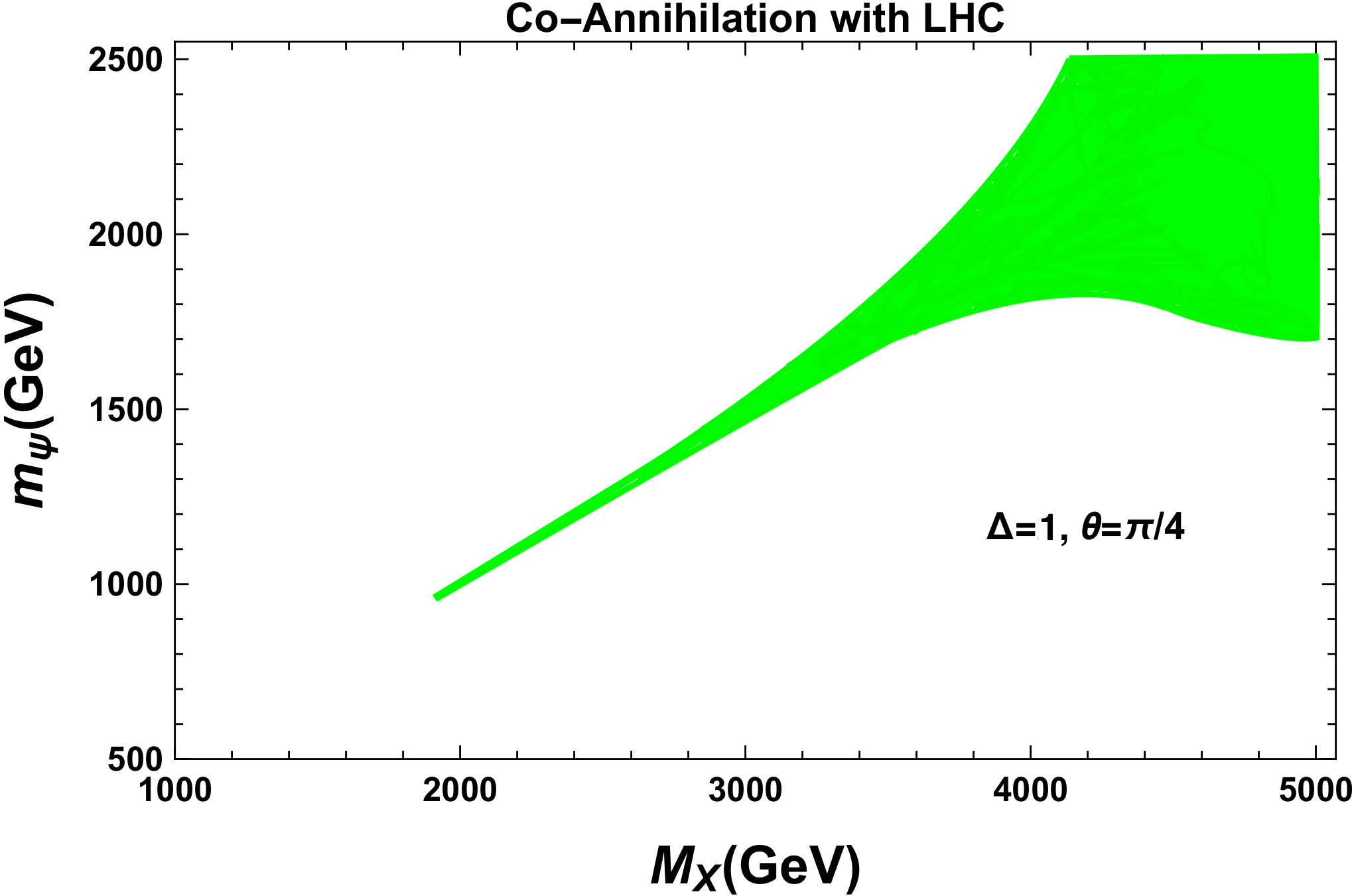}&
\includegraphics[width=7cm,height=5cm]{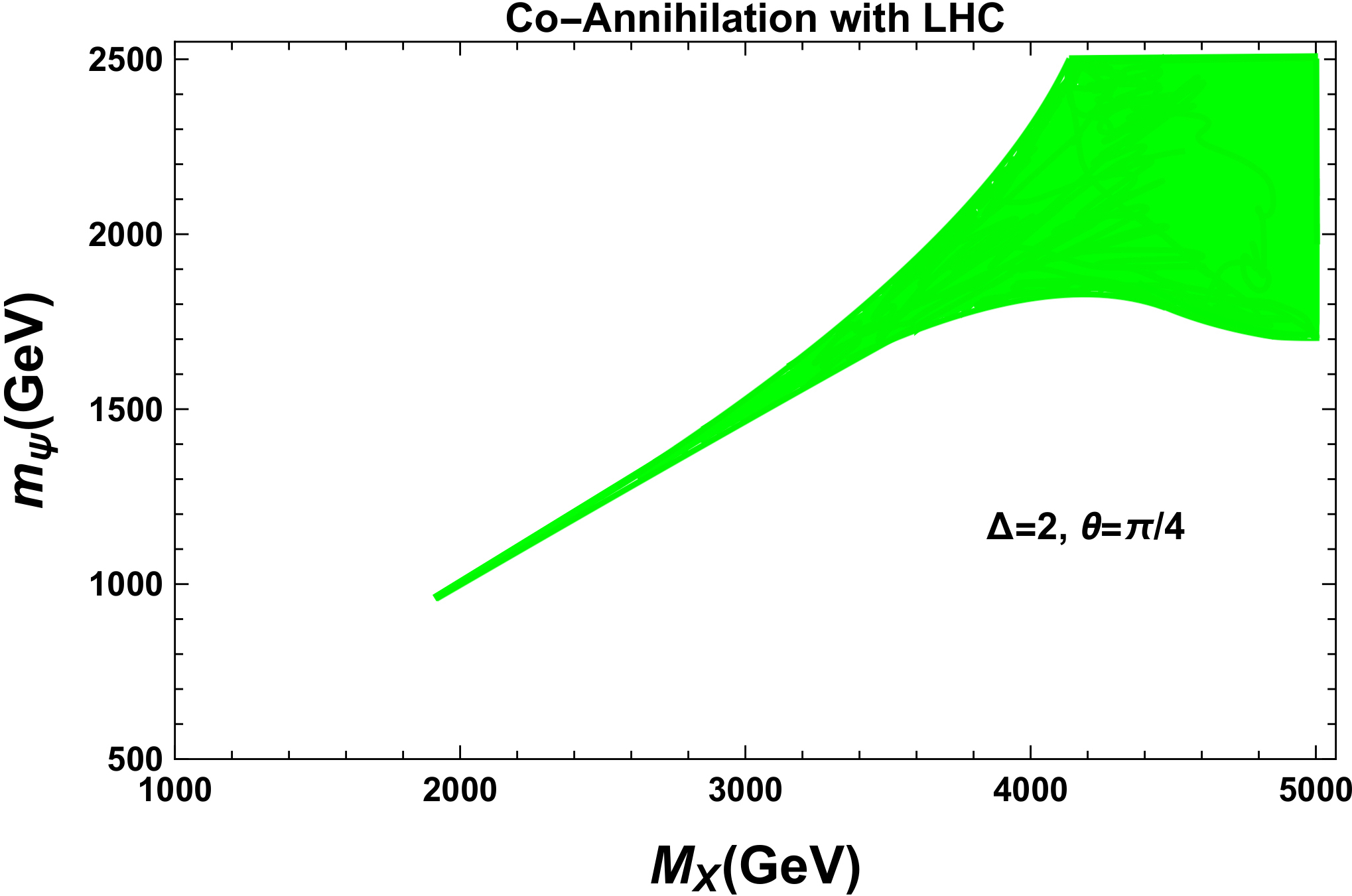}\\
(a) & (b)\\
\includegraphics[width=7cm,height=5cm]{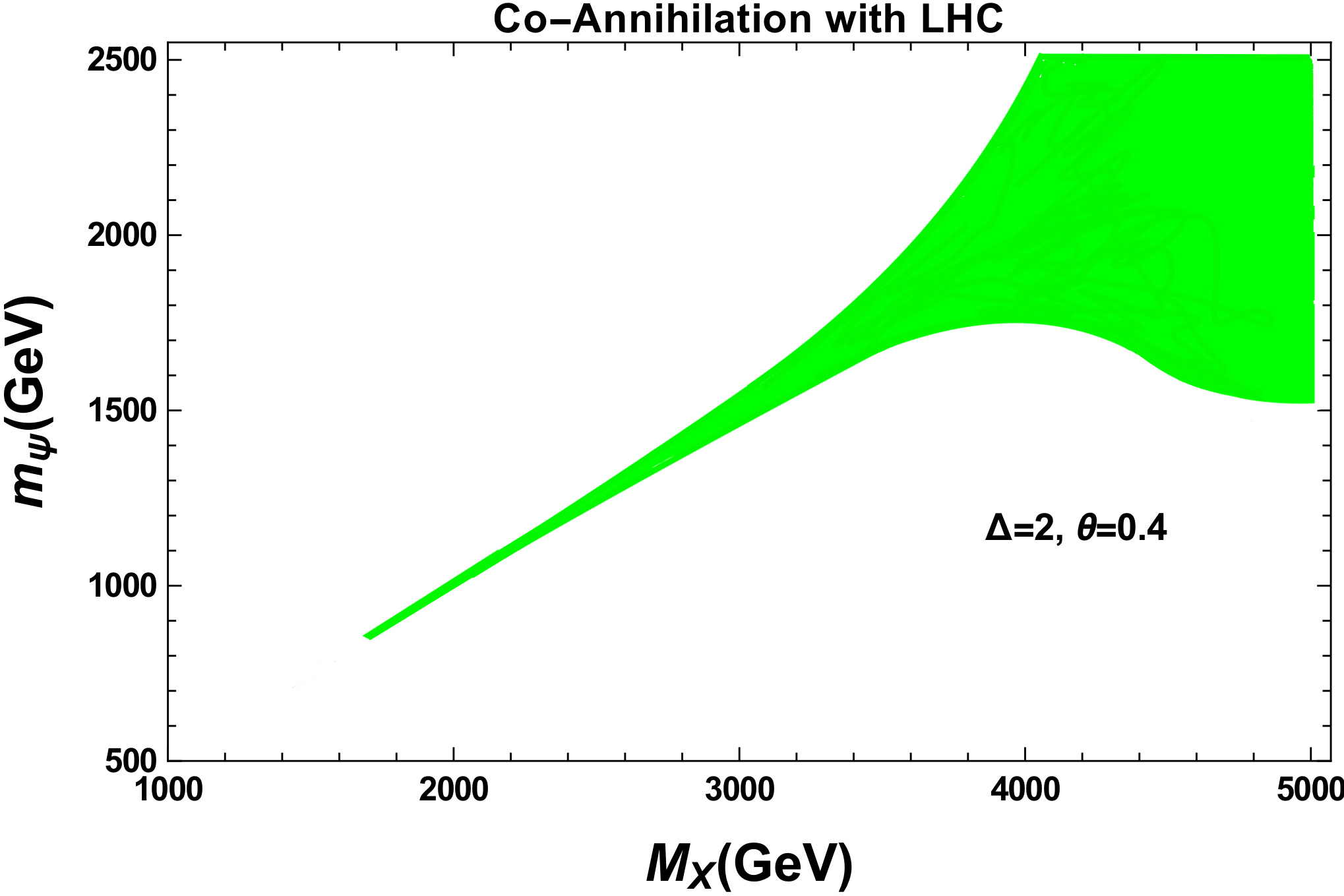}&
\includegraphics[width=7cm,height=5cm]{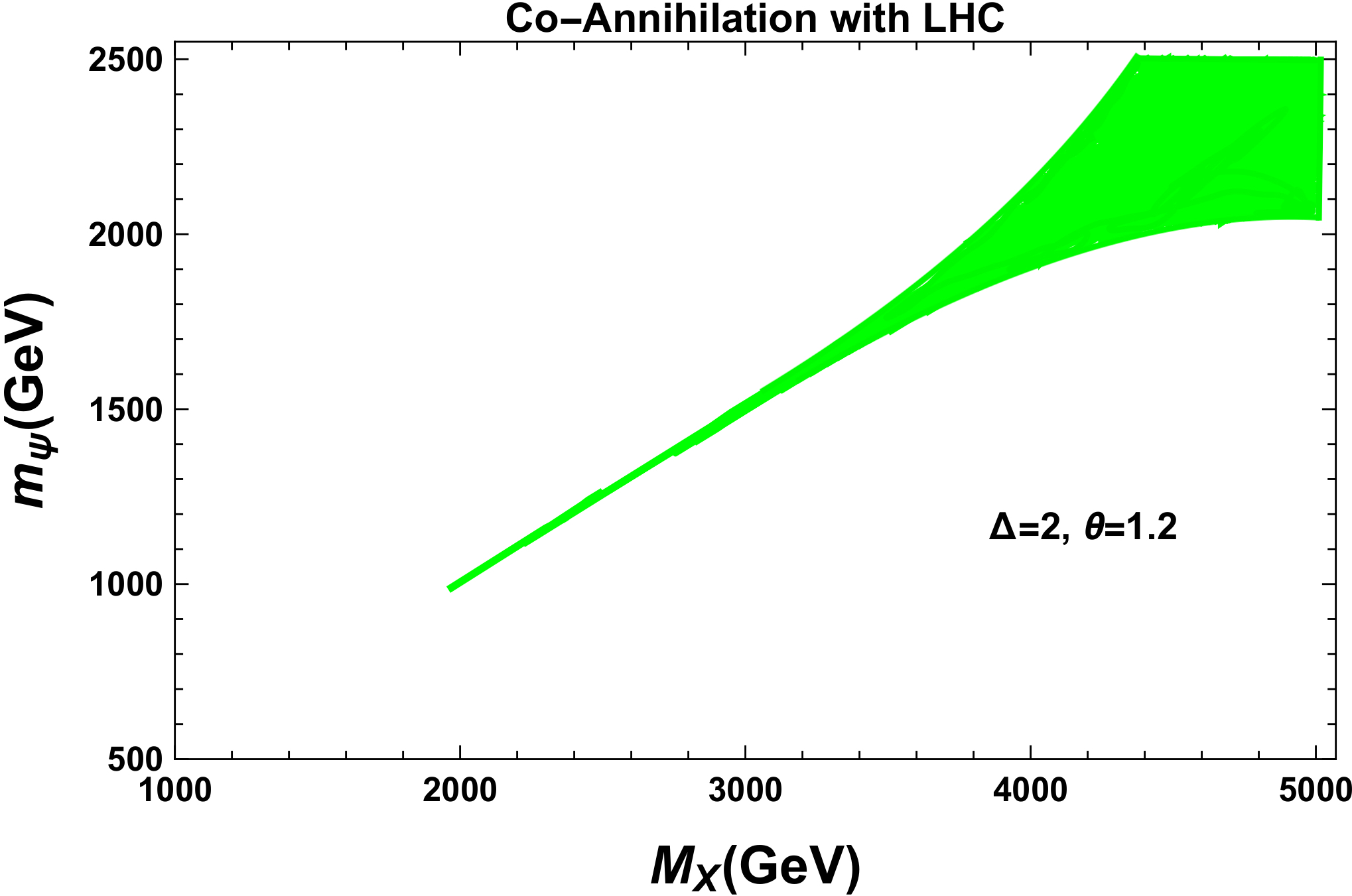}\\
(c) & (d)\\

\end{tabular}
\caption{Allowed green region in the $M_{X}$ and $m_{\psi}$ plane for co-annihilation channels with different choices of $\Delta= 2,$   $\theta = 0.4,~\pi/4,$ 1.2 and $\Delta= 1$,   $\theta = \pi/4$ (in radian). The allowed region satisfies LHC bound and PLANCK relic density bound $\Omega h^{2}\in(0.1188,~0.1212)$ .}
\label{cofig2}
\end{figure}
  
For freeze-out temperature $T \sim m_{\psi}$ 
one can see from equation \eqref{sigeff2} and \eqref{coupling} and \eqref{12} that $\sigma_{eff}$ increase with increase in $\theta$ values and the constraint on relic density in equation \eqref{omegaf} reduces the allowed parameter space of dark matter mass for any $M_X$ value  further with increase in $\theta$ values. 
In Fig. \ref{cofig1}  (c)  the chosen $\theta$ value is relatively smaller and the allowed region of $M_X$ and $m_\psi$ is more than those in (b) and (d).

Fig. \ref{cofig2} is like previous Fig. \ref{cofig1} but with LHC constraint on $M_X$ and $g_X$ imposed. Because of that, the
allowed region for lower $M_X$ is very much reduced while for higher $M_X$ above 4000 GeV there is more allowed region. This feature is similar to Fig. \ref{noco} (b).   Like Fig. \ref{cofig1} (a) and (b), the allowed region is almost same  in Fig. \ref{cofig2} (a) and (b). This shows that for $\Delta \geq 1$ with same value of $\theta$ there is insignificant change in the allowed region for the same reason as discussed in the context of previous figure. With the variation of $\theta$ again the similar feature appears like previous figure - namely for higher $\theta $ value there is lesser allowed region. In Fig. \ref{cofig2}  (c)  the chosen $\theta$ value is relatively smaller and the allowed region of $M_X$ and $m_{\psi}$ is more than those in (b) and (d). Due to LHC constraint in Fig. \ref{cofig2}, for different cases there are different lower bounds on $M_X$ and $m_{\psi}$ depending  on different set of values of $\Delta$ and $\theta$. This feature is similar to
Fig. \ref{noco} (b). For higher $\theta$ value the lower bound values for both these parameters increases to some extent with lesser allowed region. As for example,
in Fig. \ref{cofig2} (d) the lower bound on $M_X$ and $m_{\psi}$ is at about 2000 GeV and 900 GeV respectively which is relatively higher than those in Fig. \ref{cofig2} (c). 

\begin{table}[h!]
    \centering
    \begin{tabular}{| >{\centering\arraybackslash}m{0.5in} | >{\centering\arraybackslash}m{0.5in} | >{\centering\arraybackslash}m{0.5in} | >{\centering\arraybackslash}m{0.7in} |
    >{\centering\arraybackslash}m{0.7in} |
    >{\centering\arraybackslash}m{1.2in} |
    >{\centering\arraybackslash}m{1in} |}
    
\hline
$\Delta $ &$ \theta$ & $g_{X}$ & $M_{X}$(GeV)& $m_{\psi}$(GeV)& $\sqrt{u_{1}^{2}+9u_{2}^{2}}$(TeV) & $ u_{1}=u_{2}$(TeV)\\
\hline
{2} & 0.4 & 0.021 & 1702& 850&40.523& 12.814 \\ 
& $\pi/4$ & 0.035 & 1910 & 960 &27.285&8.628 \\
& 1.2 & 0.025 & 1970 & 990 &39.40&12.45 \\
\hline
{1} & $\pi/4$ & 0.035 & 1910 & 960 &27.285&8.628 \\
\hline
\hline
\multicolumn{2}{|c|}{No Co-annihilation}& 0.015 & 1645 & 825 &54.833 &17.33 \\
\hline
\end{tabular}
    \caption{Lower values of $M_{X}$ and $m_{\psi}$ and corresponding $g_{X}$ and $vev$s.}
    \label{lasttable}
\end{table}
Based on Fig. \ref{noco} (b) and Fig. \ref{cofig2} we have shown the lower values of $M_X$ and $m_{\psi}$ in the Table \ref{lasttable}. The $g_X$ value as mentioned in the table has been taken from the data file corresponding to the lower values of $M_X$ and $m_\psi$ related to these Fig.s. The {\it vev} of $\chi_1$ and $\chi_2$ fields have been evaluated using the tree level relationship of the model as given in Eq. \ref{massxboson} to get an understanding of the possible scale of extra $U(1)$ spontaneous symmetry breaking. For co-annihilation channel the {\it vev} of $\chi_1$ and $\chi_2$ field (assuming them to be equal) are around 10 TeV whereas for no co-annihilation channel it is around 17 TeV.

\section{Higher order effect on $Z-X$ mixing}
\begin{figure}[htb]
 \includegraphics[scale=0.8]{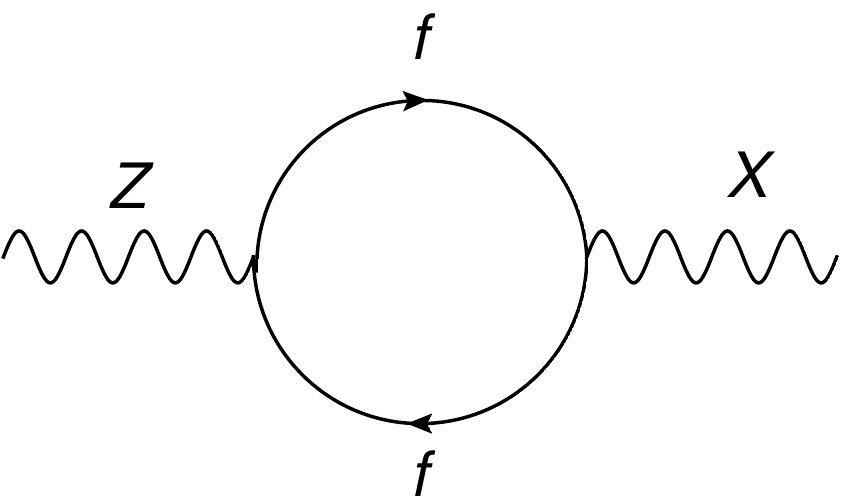}
 \caption{One-loop Feynman diagram contributing to $Z-X$ mixing}
 \label{zxloop}
 \end{figure}

Now we address the question of $Z-X$ mixing due to higher order corrections coming from one-loop Feynman diagrams as shown in fig \ref{zxloop}.  The contribution from the fermions in the loop is proportional to axial-vector coupling only and as such for our choices of $n_1 = n_4$ (corresponding to tree-level zero $Z-X$ mixing) the axial-vector coupling vanishes for quarks as shown in table III giving zero contribution to $M_{ZX}^2$ due to quarks in the loop. The dark fermion is not possible in the loop diagram as there is no coupling of $Z$ with dark fermion due to zero $Z-X$ mixing considered at the tree level.  One loop correction to $M_{ZX}^2$ will have main contribution coming from $\tau$ lepton in the loop because of non-zero axial coupling for our choice of $n_1 = n_4$.  The general expression for one loop correction to $M_{ZX}^2$  can be written as
\begin{eqnarray}
\delta M_{ZX}^2\approx -\frac{1}{4\pi^2}\sum_{f}\;c_{A}\; g_{Z}\;g_{f_a}\;{m_{f}}^2 \left(-\frac{1}{\epsilon} + \gamma_{e}-\log(4 \pi)+\log\frac{{m_f}^2}{\mu^2}\right)
	\end{eqnarray}

	Here $g_Z$ is the SM neutral coupling constant and $c_{A}$ is the coefficient depending on the particular SM fermion coupling with $Z$ boson, and $g_{f_a}$ is the axial-vector coupling of SM fermions with $X$ boson as given in table III. For completeness we give the result for one loop correction to  $M_{XX}^2$. As we do not know its tree level value from the model this correction will not be used in our numerical computation of mixing. Main contribution to it will be coming from dark fermion $\psi_1$ and $\psi_2$ and also from top quark as fermion in the loop and the  one loop correction to $  M_{XX}^2$ due to dark matters are given by  
	\begin{eqnarray}
	\delta M_{XX}^2(\psi_1)\approx-\frac{M_{XX}^2}{12\pi^2}\left(\frac{g_{11}^2}{4}-\frac{3 {m_{\psi_1}}^2}{8 M_{XX}^2}\right)\left(\frac{1}{\epsilon} - \gamma_{e}+\log(4 \pi)-\log\frac{{m_{\psi_1}}^2}{\mu^2}\right)
	\label{loopX1}
	\end{eqnarray}
	Contribution to $\delta M_{XX}^2$ due to $\psi_2$ in the fermionic loop will have similar expression like above with the replacement of $g_{11}$ by $g_{22}$ and $m_{\psi_1}$ by $m_{\psi_2}$. One loop correction to $  M_{XX}^2$ due to top quark and other fermions are given by 
	\begin{eqnarray}
	\delta M_{XX}^2({\footnotesize SM fermion})&\approx & -\frac{N_c\;M_{XX}^2}{12\pi^2}\sum_{f}\left(\frac{(g_{f_v}^2+g_{f_a}^2)}{4}-\frac{3 {m_{f}}^2}{8 M_{XX}^2}\right)\nonumber \\ && \times \left(\frac{1}{\epsilon} - \gamma_{e}+\log(4 \pi)-\log\frac{{m_f}^2}{\mu^2}\right)
	\label{loopXf}
	\end{eqnarray}
	in which $g_{f_v}$ and $g_{f_a}$ are given explicitly in terms of $n_1, n_4$ and $g_{X}$ in table III and $N_c=3$ for quarks and $N_C=1$ for leptons as well as for dark fermions.
	
	 We have followed $\overline{MS}$ scheme for numerical evaluation of $Z-X$ mixing.  The corresponding mixing angle $\theta_{ZX}$ is given by
	\begin{eqnarray}
	\tan{2\theta_{ZX}}=\frac{2\left(M_{ZX}+\delta M_{ZX}\right)}{M_{XX}-M_{ZZ}}
	\end{eqnarray}
	In $M_{ZX}$ the one loop correction is included for numerical evaluation. As an example to find a  possible value of mixing we have considered $g_{X}=0.035,~ M_{XX}=1910~\mbox{GeV}, ~m_{\psi}=960~\mbox{GeV}$  as one set of  values from Table \ref{lasttable} for lowest possible values of $M_X$ and $m_{\psi}$ for which mixing could be little bit larger and have used $M_{ZZ}$ as the experimentally measured value of $Z$ boson as $91.18~\mbox{(GeV)}$. For such choices we find $\theta_{ZX} = 6\times10^{-5}$, which is quite a small number and much lesser than the possible experimental bound \cite{mix} of about $10^{-2}$ and one may consider this mixing to be almost zero with $n_1=n_4$ even after higher order correction.

\section{Conclusion}
 Apart from Planck data constraint on relic abundance and LHC constraint on an extra $U(1)$ gauge boson mass and its gauge coupling, we have taken into account the possible constraints coming from active, sterile neutrino masses and their mixing from different oscillation experiments to find the allowed region in $M_X$ and $m_\psi$ plane. In the extra $U(1)$ gauge model considered by us, the oscillations constraints - particularly active-sterile mixing have led to the requirement of non-zero mixing $\theta$ between dark matter $\psi_1$ with the other heavy right handed Majorana fermion $\psi_2$. This has led to the consideration of co-annihilation channel for the dark matter. Also the oscillation data constrains the allowed region of $\theta$ and $\Delta$ as shown in Fig. \ref{figthetadel2}.  The allowed region in $M_X$ and $m_{\psi}$ plane is found to be reduced for co-annihilation channel with respect to no co-annihilation channel. Particularly the allowed region with co-annihilation channel is sensitive to $\theta$ value as shown in section V and for higher $\theta$ values with same $\Delta$ value there is lesser allowed region in the $M_X$ and $m_\psi$ plane. 
 
 The other important thing is that particularly with LHC constraint, in general there is some kind of lower bound on both $M_X$ and $m_{\psi}$ as shown in Fig. \ref{noco} (b)  and Fig. \ref{cofig2} in section V. Using the corresponding $g_X$ value from table \ref{lasttable} as followed from our numerical analysis for such lower values of $M_X$ and $m_{\psi}$ and using  the tree level relationship of the model as given in Eq. \ref{massxboson} which connects  $vev$s of $\chi_1$ and $\chi_2$ with mass of extra gauge boson $M_X$, we get an understanding of the possible scale of extra $U(1)$ spontaneous symmetry breaking as shown in Table \ref{lasttable}.  However, for higher values of $M_X$ such specific conclusion is difficult to obtain because of multiple possible values of $M_X$, $m_{\psi}$ and $g_X$ in the allowed region.
 The numerical analysis has been done considering zero $Z-X$ mixing at the tree level with $n_1=n_4$ which alter insignificantly even after including higher order corrections and satisfies various phenomenological low energy constraints.
With the improvement on the constraint on extra $U(1)$ gauge boson mass and its gauge coupling from LHC experiments, the allowed region in $M_X$ and $m_\psi$ plane could be further reduced and the lower bounds on $M_X$ and $m_\psi$  could be further higher.

{\bf ACKNOWLEDGMENTS}

I. A. B would like to thank Department of Science and Technology (DST), Govt. of India for financial support through Inspire Fellowship
(DST/INSPIRE Fellowship/2014/IF140038). The authors would like to thank Sushant G. Ghosh for providing HPC facilities and Puneet Sharma 
for helpful computational understandings.


\newpage
\begin{thebibliography}{150}
\bibitem{Bertone:2004pz} 
  G.~Bertone, D.~Hooper and J.~Silk,
  Phys.\ Rep.\  {\bf 405}, 279 (2005).
\bibitem{symmetry}
 J. C. Montero and V. Pleitez, Phys. Lett. B {\bf 675}, 64 (2009); 
 E. Ma and R. Srivastava, Phys. Lett. B {\bf 741}, 217 (2015); 
 S. Singirala, R. Mohanta and S. Patra, Eur.\ Phys.\ J.\ Plus {\bf 133},, 477 (2018);
 V. De Romeri, E. Fernandez-Martinez, J. Gehrlein, P. A. N. Machado and V. Niro,  J.
High Energy Phys. {\bf 10}, (2017) 169;
 E. Bertuzzo, P. A. N. Machado, Z. Tabrizi and R. Zukanovich Funchal, J.
High Energy Phys. {\bf 11},  (2017) 004; M. D. Campos, D. Cogollo, M. Lindner, T. Melo, F. S. Queiroz and W. Rodejohann, JHEP
 {\bf 08}, (2017) 092.

\bibitem{Langacker:2008yv} 
  P.~Langacker,
  Rev.\ Mod.\ Phys.\  {\bf 81}, 1199 (2009).
 \bibitem{u2}
 D. Suematsu and Y. Yamagishi, Int. J. Mod. Phys. A {\bf 10} (1995);
  M. Cvetic and S. Godfrey, [hep-ph/9504216];
  A. Leike,  Phys.Rep. {\bf 317} (1999);
  M. Cvetic and P. Langacker,  arXiv:[hep-ph/9707451];
 D. A. Demir, Phys. Rev. D {\bf 59} (1998) 015002; 
 J. L Diaz-Cruz, J. M. Hernandez-Lopez, and J. A Orduz-Ducuara, J. Phys G {\bf 40}, 125002 (2013).
\bibitem{Ma:2006km} 
  E.~Ma,
  Phys.\ Rev.\ D {\bf 73}, 077301 (2006).
  \bibitem{mix}
  J. Erler, P. Langacker, S. Munir, and E. Rojas, J. High Energy Phys. {\bf 08} (2009) 017; 
   F. del Aguila, J. de Blas, and M. Perez-Victoria, J. High Energy Phys. {\bf 09} (2010) 033; 
   R. Diener, S. Godfrey, and I. Turan, Phys.Rev. D {\bf 86}, 115017 (2012) ;
    J. Erler and P. Langacker, Phys. Lett. B {\bf 456}, 68 (1999).
  \bibitem{Zwicky:1933gu} 
  F.~Zwicky,
  Helv.\ Phys.\ Acta {\bf 6}, 110 (1933);
  [Gen.\ Relativ.\ Gravit.\  {\bf 41}, 207 (2009)].
  \bibitem{wmap}
   D. N. Spergel, et al., (WMAP Collaboration),  Astrophys. J. {\bf 148} (2003) 175;  M. Tegmark, et al., (SDSS
Collaboration ), Phys. Rev. D {\bf{69}}  103501 (2004).
\bibitem{Aghanim:2018eyx} 
  N.~Aghanim {\it et al.} (Planck Collaboration),
  arXiv:1807.06209.
  \bibitem{Aaboud:2017yvp} 
  M.~Aaboud {\it et al.} (ATLAS Collaboration),
  Phys.\ Rev.\ D {\bf 96}, 052004 (2017).
\bibitem{Aaboud:2017buh} 
  M.~Aaboud {\it et al.} (ATLAS Collaboration),
  J. High Energy Phys. {\bf 10},  (2017) 182.
\bibitem{cms} 
  A.~M.~Sirunyan {\it et al.} (CMS Collaboration),
  Phys.\ Lett.\ B {\bf 769}, 520 (2017);{\bf 772}, 882 {\bf (E)} (2017)].
 \bibitem{Gonzalez-Garcia:2015qrr} 
  M.~C.~Gonzalez-Garcia, M.~Maltoni, and T.~Schwetz,
  Nucl.\ Phys. {\bf B908}, 199 (2016).
\bibitem{Ahn:2006zza} 
  M.~H.~Ahn {\it et al.} (K2K Collaboration),
  Phys.\ Rev.\ D {\bf 74}, 072003 (2006).

\bibitem{Adamson:2017zcg} 
  P.~Adamson {\it et al.} (NOvA Collaboration),
  Phys.\ Rev.\ D {\bf 96},  072006 (2017).

\bibitem{fermi}
 A. A. Aguilar-Arevalo,  Phys.\ Rev.\ Lett.\  {\bf 121}, 221801 (2018);
   A.~M.~Sirunyan {\it et al.} (CMS Collaboration),
  J. High Energy Phys. {\bf 08},  (2018) 130; 06 (2018) 120.
  \bibitem{Adhikari:2015woo} 

R.~Adhikari, D.~Borah, and E.~Ma,
  Phys.\ Lett.\ B {\bf 755}, 414 (2016).
  
  \bibitem{okada}
  N.~Okada, S.~Okada, and D.~Raut,  Phys.\ Rev.\ D {\bf 100},  035022 (2019); 
  T. Nomura and H. Okada, Eur. Phys. J. C {\bf 78},  189 (2018); 
  N.~Okada and S.~Okada,Phys.\ Rev.\ D {\bf 95}, 035025 (2017); 
   S.~Singirala, R.~Mohanta, S.~Patra, and S.~Rao,
  J. Cosmol. Astropart. Phys. {\bf 11}, (2018) 026;
   Z.~L.~Han and W.~Wang,
  Eur.\ Phys.\ J.\ C {\bf 78}, 839 (2018);
  M.~Escudero, S.~J.~Witte, and N.~Rius,
  J. High Energy Phys. {\bf 08}, (2018) 190.
  \bibitem{ma1}
  E. Ma and D. Suematsu, Mod. Phys. Lett. A {\bf 24}, 583 (2009).
  \bibitem{gfit} M.~Dentler, Á.~Hernández-Cabezudo, J.~Kopp, P.~A.~N.~Machado, M.~Maltoni, I.~Martinez-Soler, and T.~Schwetz,
  J. High Energy Phys. {\bf 08},  (2018) 010.
\bibitem{Vagnozzi:2018jhn} 
  S.~Vagnozzi {\it et al},
  Phys.\ Rev.\ D {\bf 98}, 083501 (2018).
\bibitem{Choudhury:2018byy} 
  S.~Roy Choudhury and S.~Choubey,
  J. Cosmol. Astropart. Phys. {\bf 09}, (2018) 017.

\bibitem{Beltran:2008xg} 
  M.~Beltran, D.~Hooper, E.~W.~Kolb, and Z.~C.~Krusberg,
  Phys.\ Rev.\ D {\bf 80}, 043509 (2009).
  \bibitem{Gondolo:1990dk} 
  P.~Gondolo and G.~Gelmini,
  Nucl.\ Phys. {\bf B360}, 145 (1991).
  \bibitem{Kolb:1990vq} 
  E.~W.~Kolb and M.~S.~Turner, {\it The Early Universe. Frontiers in Physics}
   (Avalon Publishing, Redwood City, 1994).
\bibitem{Griest} 
  K.~Griest and D.~Seckel,
  Phys.\ Rev.\ D {\bf 43}, 3191 (1991).
\bibitem{Cannoni:2015wba} 
  M.~Cannoni,
  Eur.\ Phys.\ J.\ C {\bf 76}, 137 (2016).
\bibitem{Berlin:2014tja} 
  A.~Berlin, D.~Hooper, and S.~D.~McDermott,
  Phys.\ Rev.\ D {\bf 89}, 115022 (2014).

\end{thebibliography}
	\end{document}